\documentclass[fleqn,usenatbib]{mnras}
\include{notations}
\usepackage{slashed}
\usepackage{graphicx}
\usepackage{grffile}
\usepackage[normalem]{ulem}
\usepackage{dcolumn}
\usepackage{xcolor}
\usepackage{booktabs}
\usepackage{amsmath}
\usepackage{amssymb}
\usepackage{slashed}
\usepackage{todonotes}
\usepackage{comment}
\usepackage{subfigure}
\usepackage{lastpage}
\usepackage{xspace}

\newcolumntype{M}{R@{${}\pm{}$}L}

\usepackage{multirow}
\usepackage{arydshln}
\usepackage[T1]{fontenc}
\usepackage{ae,aecompl}
\usepackage{lineno}
\usepackage[export]{adjustbox}

\usepackage{txfonts}

\usepackage{eso-pic}

\AddToShipoutPictureBG*{%
  \AtPageUpperLeft{%
    \hspace{0.75\paperwidth}%
    \raisebox{-3.5\baselineskip}{%
      \makebox[0pt][l]{\textnormal{DES-2019-0483}}
}}}%

\AddToShipoutPictureBG*{%
  \AtPageUpperLeft{%
    \hspace{0.75\paperwidth}%
    \raisebox{-4.5\baselineskip}{%
      \makebox[0pt][l]{\textnormal{FERMILAB-PUB-22-707-PPD}}
}}}%

\newcommand{\red}[1]{{\textcolor{red}{#1}}}

\newcommand{\redmapper}{\textit{redMaPPer}\xspace}
\newcommand{\redmagic}{\textit{redMaGiC}\xspace}

\newcommand{\lcdm}{$\Lambda$CDM\xspace}
\newcommand{\wcdm}{$w$CDM\xspace}
\newcommand{\gammat}{\ensuremath{\gamma_{t}(\theta)}\xspace}
\newcommand{\wtheta}{\ensuremath{w(\theta)}\xspace}
\newcommand{\xipm}{\ensuremath{\xi_{\pm}(\theta)}\xspace}
\newcommand{\Om}{\ensuremath{\Omega_{\rm m}}\xspace}
\newcommand{\vecn}{\boldsymbol{\hat{\textbf{n}}}}
\newcommand{\vecl}{\vec{l}}

\newcommand{\deltagobsi}[1]{\delta^{#1}_{g,\text{obs}}}
\newcommand{\twobytwo}{2$\times$2pt}
\newcommand{\threebytwo}{3$\times$2pt}

\newcommand{\maglim}{\textit{MagLim}\xspace}

\newcommand{\code}[1]{\texttt{#1}}
\newcommand{\balrog}{\code{Balrog}\xspace}
\newcommand{\Balrog}{\code{BALROG}\xspace}
\newcommand{\dx}[1]{\mathrm{d}{#1}\,}

\newcommand{\csample}{\ensuremath{C_{\rm sample}}\xspace}
\newcommand{\ctotal}{\ensuremath{C_{\rm total}}\xspace}
\newcommand{\cbalrog}{\ensuremath{C^{\balrog}_{\rm sample}}\xspace}
\newcommand{\cdata}{\ensuremath{C^{\rm Data}_{\rm sample}}\xspace}
\newcommand{\cnbody}{\ensuremath{C^{\rm N-body}_{\rm sample}}\xspace}
\newcommand{\carea}{\ensuremath{C_{\text{area}}}\xspace}
\newcommand{\buzzard}{\textsc{Buzzard}\xspace}
\newcommand\eqn[1]{equation~\ref{#1}}

\newcommand{\ec}[1]{Eq.~(\ref{eq:#1})}

\newcommand{\eql}[1]{\label{eq:#1}}

\newcommand{\No}{\ensuremath{N(0)}}
\newcommand{\Nk}{\ensuremath{N(\delta\kappa)}}
\newcommand{\Nko}{\ensuremath{N(\delta\kappa + 0)}}
\newcommand{\Noonly}{\ensuremath{N(0 \text{ only})}}
\newcommand{\Nkonly}{\ensuremath{N(\delta\kappa \text{ only})}}
\newcommand{\dk}{\ensuremath{\delta\kappa}}

\newcommand\fig[1]{Figure~\ref{#1}}

\newcommand\sect[1]{Section~\ref{#1}}
\newcommand\tab[1]{Table~\ref{#1}}

\newcommand{\NM}[1]{{\color{purple}[NM: #1]}}

\title[Dark Energy Survey Year 3 results: Magnification]{Dark Energy Survey Year 3 results:  Magnification modeling and impact on cosmological constraints from galaxy clustering and galaxy-galaxy lensing}

\date{\today}

\pubyear{2022}

\author[Elvin-Poole \& MacCrann et al.]{
\parbox{\textwidth}{
\Large
J.~Elvin-Poole,$^{1,2,3,*}$
N.~MacCrann,$^{4,\dag}$
S.~Everett,$^{5}$
J.~Prat,$^{6,7}$
E.~S.~Rykoff,$^{8,9}$
J.~De~Vicente,$^{10}$
B.~Yanny,$^{11}$
K.~Herner,$^{11}$
A.~Fert\'e,$^{5}$
E.~Di Valentino,$^{12}$
A.~Choi,$^{13}$
D.~L.~Burke,$^{8,9}$
I.~Sevilla-Noarbe,$^{10}$
A.~Alarcon,$^{14}$
O.~Alves,$^{15}$
A.~Amon,$^{16,17}$
F.~Andrade-Oliveira,$^{15}$
E.~Baxter,$^{18}$
K.~Bechtol,$^{19}$
M.~R.~Becker,$^{14}$
G.~M.~Bernstein,$^{20}$
J.~Blazek,$^{21}$
H.~Camacho,$^{22,23}$
A.~Campos,$^{24}$
A.~Carnero~Rosell,$^{25,23,26}$
M.~Carrasco~Kind,$^{27,28}$
R.~Cawthon,$^{29}$
C.~Chang,$^{6,7}$
R.~Chen,$^{30}$
J.~Cordero,$^{12}$
M.~Crocce,$^{31,32}$
C.~Davis,$^{8}$
J.~DeRose,$^{33}$
H.~T.~Diehl,$^{11}$
S.~Dodelson,$^{24,34}$
C.~Doux,$^{35}$
A.~Drlica-Wagner,$^{6,11,7}$
K.~Eckert,$^{20}$
T.~F.~Eifler,$^{36,5}$
F.~Elsner,$^{37}$
X.~Fang,$^{38,36}$
P.~Fosalba,$^{31,32}$
O.~Friedrich,$^{17}$
M.~Gatti,$^{20}$
G.~Giannini,$^{39}$
D.~Gruen,$^{40}$
R.~A.~Gruendl,$^{27,28}$
I.~Harrison,$^{41}$
W.~G.~Hartley,$^{42}$
H.~Huang,$^{36,43}$
E.~M.~Huff,$^{5}$
D.~Huterer,$^{15}$
E.~Krause,$^{36}$
N.~Kuropatkin,$^{11}$
P.-F.~Leget,$^{8}$
P.~Lemos,$^{37,44}$
A.~R.~Liddle,$^{45}$
J.~McCullough,$^{8}$
J.~Muir,$^{46}$
J.~Myles,$^{47,8,9}$
A. Navarro-Alsina,$^{48}$
S.~Pandey,$^{20}$
Y.~Park,$^{49}$
A.~Porredon,$^{1,2,50}$
M.~Raveri,$^{35,20}$
M.~Rodriguez-Monroy,$^{35,10}$
R.~P.~Rollins,$^{12}$
A.~Roodman,$^{8,9}$
R.~Rosenfeld,$^{51,23}$
A.~J.~Ross,$^{1}$
C.~S{\'a}nchez,$^{20}$
J.~Sanchez,$^{11}$
L.~F.~Secco,$^{7}$
E.~Sheldon,$^{52}$
T.~Shin,$^{53}$
M.~A.~Troxel,$^{30}$
I.~Tutusaus,$^{54,31,32}$
T.~N.~Varga,$^{55,56,57}$
N.~Weaverdyck,$^{15,33}$
R.~H.~Wechsler,$^{47,8,9}$
B.~Yin,$^{24}$
Y.~Zhang,$^{58}$
J.~Zuntz,$^{50}$
M.~Aguena,$^{23}$
S.~Avila,$^{35,59}$
D.~Bacon,$^{60}$
E.~Bertin,$^{61,62}$
S.~Bocquet,$^{40}$
D.~Brooks,$^{37}$
J.~Garc\'ia-Bellido,$^{59}$
K.~Honscheid,$^{1,2}$
M.~Jarvis,$^{20}$
T.~S.~Li,$^{63}$
J. Mena-Fern{\'a}ndez,$^{10}$
C.~To,$^{1}$
and R.D.~Wilkinson$^{44}$
\begin{center} (DES Collaboration) \end{center}
\emph{\normalsize Affiliations are listed at the end of the paper}
}
\vspace{0.4cm}
\\
}

\begin{document}
\label{firstpage}
\pagerange{\pageref{firstpage}--\pageref{lastpage}}
\maketitle

\begin{abstract}
We  study the effect of magnification in the Dark Energy Survey Year 3 analysis of galaxy clustering and galaxy-galaxy lensing, using two different lens samples: a sample of Luminous red galaxies, \redmagic, and a sample with a redshift-dependent magnitude limit, \maglim. We account for the effect of magnification on both the flux and size selection of galaxies, accounting for systematic effects using the \balrog image simulations. 
We estimate the impact of magnification  on the galaxy clustering and galaxy-galaxy lensing cosmology analysis, finding it to be a significant systematic for the \maglim sample. 
We show cosmological constraints from the galaxy clustering auto-correlation and galaxy-galaxy lensing signal with different magnifications priors, finding broad consistency in cosmological parameters in \lcdm and \wcdm. However, when magnification bias amplitude is allowed to be free, we find the two-point correlations functions prefer a different amplitude to the fiducial input derived from the image simulations. We validate the magnification analysis by comparing the cross-clustering between lens bins with the prediction from the baseline analysis, which uses only the auto-correlation of the lens bins, indicating systematics other than magnification may be the cause of the discrepancy. We show adding the cross-clustering between lens redshift bins to the fit significantly improves the constraints on lens magnification parameters and allows uninformative priors to be used on magnification coefficients, without any loss of constraining power or prior volume concerns.

\end{abstract}


\makeatletter
\def \blfootnote{\xdef\@thefnmark{}\@footnotetext}
\makeatother
\blfootnote{$^{\star}$ E-mail: jack.elvinpoole@gmail.com}
\blfootnote{$^{\dag}$ E-mail: nm746@cam.ac.uk}

\section{Introduction}
\label{sec:intro}

Although astronomers have a long history of mapping out the projected distribution of galaxies on the sky, cosmological models make the cleanest predictions about the three-dimensional distribution of mass in the universe, i.e.\ the dark-matter dominated, total matter distribution. The relation between galaxy and matter density, known as the galaxy bias, is difficult to predict theoretically, hence it is difficult to extract cosmological information from maps of projected galaxy density alone.
Gravitational lensing provides a relatively direct way to probe the total mass distribution that galaxies sit within. In particular, the mass associated with foreground, \emph{lens}, galaxies distorts the observed shapes of background, \emph{source}, allowing inference of the mass distribution around the foreground lenses, a phenomenon known as galaxy-galaxy lensing. Galaxy-galaxy lensing then can be used to break the degeneracy between the galaxy bias and the amplitude of total matter clustering, that is present in galaxy clustering measurements, and thus infer useful cosmological constraints (see e.g.\  \citealt{hujain2004, bernstein09,joachimi2010,yooseljak12,mandelbaum2013}).

The Dark Energy Survey (DES) is one of several galaxy imaging surveys aiming to exploit the combination of clustering and lensing information, with large sky area and deep imaging now returning high signal-to-noise measurements of the angular correlation function of galaxies, $w(\theta)$, and  the mean tangential shear induced in the background galaxies by the foreground lenses, $\gamma_t(\theta)$. As statistical power continues to increase, more subtle effects need to be included in the modelling of the signal. Here we focus on gravitational lensing {\it magnification}, which impacts the number of galaxies observed in a given area of sky, leading to observable effects on the galaxy clustering and galaxy-galaxy lensing statistics. Therefore, they need to be accounted for and the data from DES Year 3 data (Y3, from the first three years of DES observations) affords an excellent opportunity to detect this effect. Needless to say, as surveys get wider and deeper, this effect will become more and more important, so we view this paper as one in a series of communal attempts to incorporate magnification into cosmological analyses.
Lensing magnification has been investigated in the context of the weak lensing cosmology analyses, most recently in \citet{lorenz_2018,deshpande_2020,thiele_2020,vonWietersheim-Kramsta:2021lid, mahoney22, Duncan_2022} and has been detected in a number of different ways dating back to at least \citet{Scranton:2005ci} (and see references therein for even earlier detections).

We begin by overviewing the relevant theory in \sect{sec:theory} and we describe the data and simulations used in Sections~\ref{sec:data} \&~\ref{sec:simulations}. In \sect{sec:method} we estimate the amplitudes of the magnification contributions to our theory predictions, for both lens samples using several methods and then in \sect{sec:simresults} propagate that to a projection of what should be expected in DES Y3. In \sect{sec:buzzard_validation}, we validate
our modelling framework on cosmological simulations, and then present our results for the DES Y3 data in \sect{sec:results}.

This paper is one of three from the DES Y3 analysis presenting cosmology results from the combination of galaxy clustering and galaxy-galaxy lensing, which we will refer to as ``2$\times$2pt''. The other two are \cite{y3-2x2ptaltlensresults} which presents results from the \maglim lens sample and \cite{y3-2x2ptbiasmodelling} which presents results from the \redmagic lens sample. These results are combined with the lensing shear auto-correlation (known as \emph{cosmic shear}, see \cite{y3-cosmicshear1} and \cite*{y3-cosmicshear2}) in the the 3$\times$2pt paper \citep{y3-3x2ptkp}. 
\section{Theory}
\label{sec:theory}
In photometric surveys such as DES, we use photometric redshift estimates  to place galaxies into redshift bins, for which we have estimates of the ensemble redshift distribution.
In the absence of magnification, the intrinsic projected galaxy density contrast in redshift bin $i$, $\delta^i_{g,\mathrm{int}} (\vecn)$, is given by the line-of-sight integral of the three-dimensional galaxy density contrast 
\begin{equation}
\delta^i_{g,\mathrm{int}} (\vecn)  \simeq \int\! d\chi\, 
 W^i_{g}(\chi)
 \delta_{g}^{\rm 3D}\!\left(\vecn \chi, \chi\right)
 \end{equation}
with $\chi$ the comoving distance and $W_{g}^i = n_g^i(z)\, d z/d\chi$ the normalized selection function of galaxies in redshift bin $i$. The approximate equality acknowledges that this neglects redshift space distortions (RSD, \citealt{10.1093/mnras/227.1.1}). These are included in our modeling but suppressed here for simplicity.

In this section, we derive the modulation of the observed projected galaxy density by magnification and calculate the magnification contribution to angular two-point statistics.

\subsection{Magnification}
\label{sec:magnification_theory}

We can express magnification in terms of the convergence $\kappa$ and the shear $\gamma$ (e.g. \citealt{bartelmann01}):
\begin{equation}\label{eq:magnification_kappa}
    \mu 
    = \frac{1}{(1-\kappa)^2 - \vert \gamma \vert ^2 } \approx \frac{1}{1-2\kappa}\approx 1+ 2\kappa,
\end{equation}
in the limit of weak lensing when $\kappa \ll 1$ and $\gamma \ll 1$. 

Magnification alters the trajectory of photons such that, in regions of positive (negative) convergence (i) the apparent distance between any two points on a source plane is increased (decreased), and (ii) the telescope captures a greater (smaller) fraction of the solid angle of light emitted from an object. Magnification then impacts both the apparent position of galaxies and the distribution of light received from an individual galaxy image. In large scale structure surveys, where we are interested specifically in the observed number density of objects, the impact of magnification can be separated into the following two effects:
\begin{itemize}
    \item \textbf{Change in observed area element}: Since  the distance between the centroids of galaxy images will increase with positive convergence, this will appear to an observer as a given area element $\Delta\Omega$ on the unlensed sky being mapped to an area element of area  $\mu\Delta\Omega$ in the presence of magnification $\mu$. Hence the observed area number density of galaxies  decreases by a factor $\mu$. 
    \item \textbf{Change in selection probability of individual galaxies}: A lensing magnification $\mu$ increases the apparent distance between points within the image of the galaxy, enlarging the apparent image size, while the increased solid angle captured by the telescope increases the total flux received (such that galaxy surface brightness is conserved). To first order this increases the total observed flux by a factor $\mu$. 
    Galaxies entering a given photometric sample are selected based on their measured (i.e. observed) properties, for example their flux or size. 
    We note that in real data, galaxy selection can be complex in detail (i.e. not simply a threshold in total galaxy flux), and so accurately predicting the response of the number density to a change in magnification requires simulations of the selection function.
\end{itemize}

The overdensity due to convergence $\kappa$ at position $\vecn$ on the sky, can be written in terms of the observed galaxy number densities, $n^{\textrm{sel}}(\vecn,\kappa)$, and the same quantity at $\kappa=0$ (e.g. \citealt{joachimi2010, bernstein09}),
\begin{equation}
    \delta_g^{\mathrm{mag}}(\vecn, \kappa) = \frac{n^{\mathrm{sel}}(\vecn,\kappa)}{n^{\mathrm{sel}}(\vecn,0)} - 1.
    \label{eq:delta_g_mag}
\end{equation}
Here the superscript `sel' indicates that a selection has been applied using thresholds on various observed (i.e.\ lensed) properties of the galaxies, which we will denote by a vector $\vec{F}'$. The observed number density at position $\vecn$ can be written as an integral over $N(\vec{F},\vecn)$, the absolute number of galaxies in direction $\vecn$ with \emph{unlensed} properties $\vec{F}$, divided by the area element $\Delta\Omega(\kappa)$ (on the lensed sky)
\begin{equation}
n^{\mathrm{sel}}(\vecn,\kappa)
= \frac{1}{\Delta\Omega(\kappa)} \int \dx{\vec{F}} S(\vec{F}') N(\vec{F},\vecn)
\end{equation}
where $S(\vec{F}')$ is the sample selection function, which operates on lensed properties $\vec{F}'$. For small convergence $\kappa$, we can make the substitution $\Delta\Omega(\kappa) = \Delta\Omega(0)/(1-2\kappa)$, such that
\begin{equation}
n^{\mathrm{sel}}(\vecn,\kappa)
= \frac{1-2\kappa}{\Delta\Omega(0)} \int \dx{\vec{F}} S(\vec{F}') N(\vec{F},\vecn)
\end{equation}
We can then Taylor expand $S(\vec{F}')$ around $\kappa=0$
\begin{equation}
n^{\mathrm{sel}}(\vecn,\kappa)
\approx \frac{1-2\kappa}{\Delta\Omega(0)} \int \dx{\vec{F}} \left[S(\vec{F})+\kappa \frac{\partial{S}}{\partial{\kappa}}\right] N(\vec{F},\vecn)
\end{equation}
and drop terms involving $\kappa^2$, leading to 
\begin{align}
n^{\mathrm{sel}}(\vecn,\kappa)
&\approx \frac{1-2\kappa}{\Delta\Omega(0)} \int \dx{\vec{F}} S(\vec{F})N(\vec{F},\vecn) 
+ \kappa \frac{1}{\Delta\Omega(0)}\int \dx{\vec{F}} \frac{\partial{S}}{\partial{\kappa}} N(\vec{F},\vecn)\\
&\approx (1-2\kappa) n^{\mathrm{sel}}(\vecn,0) +  \frac{\kappa}{\Delta\Omega(0)}\int \dx{\vec{F}} \frac{\partial{S}}{\partial{\kappa}}N(\vec{F},\vecn).
\label{eq:nsel}
\end{align}

Substituting this into \eqn{eq:delta_g_mag}, we have
\begin{align} \label{eq:delta_g_full}
    \delta_g^{\mathrm{mag}}(\vecn) &= \kappa(\vecn) \left(-2 + \frac{1}{n^{\mathrm{sel}}(\vecn,0)\Delta\Omega(0)}\int \dx{\vec{F}} \frac{\partial{S}}{\partial{\kappa}}N(\vec{F},\vecn)\right)\\
    &= \kappa(\vecn) \left[-2 + \frac{1}{N^{\mathrm{sel}}(\vecn,0)}
    \frac{\partial N^{\mathrm{sel}}(\vec{\theta},0)}{\partial \kappa} \right]
    \label{eq:delta_g_mag2}
\end{align}
where
\begin{equation}
    N^{\mathrm{sel}}(\vecn,0) = \int \dx{\vec{F}} S(\vec{F})N(\vec{F},\vecn).
\end{equation}
In \eqn{eq:nsel} we can identify the first term as being the number density in the unlensed case, $n^{\mathrm{sel}}(\vecn,0)$, modulated by $(1-2\kappa)$ due to the change in area element. The second term is proportional to $\kappa$, with constant of proportionality given by the response of the number of selected objects per (unlensed) area element, to a change in $\kappa$. We can thus summarize the effect on the projected number density contrast as, 
\begin{equation}
    \delta_g^{\mathrm{mag}}(\vecn) =  \kappa(\vecn)\left[ \carea + \csample^i \right],
    \label{eq:delta_g_mag_ctotal}
\end{equation}
with $\carea=-2$, and the total magnification contribution described by a single constant $C^i=\carea + \csample^i$.

Where the galaxy selection function is simply made via a cut in magnitude, $m_{\textrm{cut}}$, this expression becomes (\citealt{joachimi2010, sv-magnification})
\begin{equation}\label{eq:delta_mag_alpha}
   \delta^i_{g,{\text{mag}}}(\vecn) = 2[\alpha^i(m_{\text{cut}})-1]\kappa^i (\vecn),
\end{equation}
where 
\begin{equation}
    \alpha^i(m) = 2.5\frac{d}{dm}[\log{N_{\mu}(m)}]
\end{equation}
and $N_{\mu}(m)$ is the (lensed) cumulative number of galaxies as a function of maximum magnitude $m$.

In this case, whether an excess magnification increases or decreases the observed number density i.e. whether the increase in observed flux wins over the dilution due to change in area element, depends on the intrinsic slope of the cumulative flux distribution. The larger the ratio of faint to bright objects in the sample, the more dominant the former effect is.

Since real galaxy samples are a complex selection of flux, color, position and shape, we estimate the response constant $\csample$ in DES Y3 using the image simulation \emph{Balrog} \citep{y3-balrog}, as described in \sect{sec:balrog}.

\subsection{Lens magnification}

The primary effect we will study is magnification of the lens sample\footnote{See Appendix~\ref{app:source_mag} for discussion of the magnification of the source sample.} by structure that is between the lenses and the observer,
\begin{equation}
    \delta^i_{g,\text{obs}} =  \delta^i_{g,\text{int}} + \delta^i_{g,\text{mag}}.
\end{equation}

Following \sect{sec:magnification_theory}, we can write the change in number density produced by magnification as proportional to the convergence, and we define $C^i$ as: 
\begin{equation}\label{eq:definition_C}
\delta^i_{g,\text{mag}} (\vecl) = C^i \kappa^i (\vecl).
\end{equation}
where $\kappa$ here denotes the convergence experienced by the lens galaxies in redshift bin $i$, and note we are now working with harmonic transform of the density contrast, $\delta^i_{g,\text{mag}} (\vecl)$. Recall that these are the galaxies whose clustering we are measuring, but we will also be cross-correlating this sample with the background source sample in the galaxy-galaxy lensing probe. The galaxies in the source sample also experience magnification, which we can ignore here since it impacts the two-point functions at higher order. See Appendix~\ref{app:source_mag} for more details on source magnification in this sample and \citealt{y3-gglensing} and \citealt{Duncan_2022} for further studies justifying the exclusion of source magnification from \twobytwo \ analyses.

Then, this change in the density contrast affects the galaxy overdensity angular power spectrum, $C_{gg}(l)$ as:
\begin{equation}
\left< \deltagobsi{i} \deltagobsi{j} \right> = \left< \delta^i_{g,\text{int}} \delta^j_{g,\text{int}} \right>  + C^iC^j\left< \kappa^i_l\kappa^i_l \right> + 2C^i \left< \delta^i_{g,\text{int}} \kappa^i_l \right> .
\label{eq:gg}
\end{equation}
where angle brackets $\left<\right>$ denote an angular power spectrum and we have dropped the $(\vecl)$ arguments for brevity.

Lens magnification also impacts galaxy-galaxy lensing since the convergence experienced by the lens galaxies is correlated with that causing the shear of the source galaxies (denoted here as $\gamma_G$), as well as their intrinsic alignment (denoted here as $\gamma_{IA}$). The angular cross-correlation power spectrum between lens galaxy overdensity of redshift bin $i$ and shape of galaxies in redshift bin $j$ is then 
\begin{align}
\eql{ggl}
\left< \deltagobsi{i} \gamma^j \right> =& \left< \delta^i_{g,\text{int}} \left(\gamma^j_{\mathrm G}+\gamma^j_{\mathrm IA} \right)\right> + C^i\left< \kappa^i_l\left(\gamma^j_{\mathrm G}+\gamma^j_{\mathrm IA} \right)\right> \\
\nonumber= &\left< \delta_g^\text{int}\left(\gamma^j_{\mathrm G}+\gamma^j_{\mathrm IA} \right) \right> + C^i\left< \kappa^i_l\gamma_G^j \right>+ C^i\left< \kappa^i_l\gamma^j_{\mathrm IA}\right>.
\end{align}


\subsection{Modeling the correlation functions}

The modeling of the two point functions is described in detail in \cite{y3-generalmethods}, here we summarize the basic structure of this computation. 

We use the Limber approximation to calculate each term contributing to the galaxy-galaxy lensing power spectrum. For two general fields, this is simply 
 \begin{align}
     \label{eq:cl}
     C_{AB}^{ij}(\ell) = \int d\chi \frac{W_A^i(\chi)W_B^j(\chi)}{\chi^2}
     P_{AB}\left(k = \frac{\ell+0.5}{\chi},z(\chi)\right)\,
 \end{align}
 where the window functions for galaxy density and shear are defined in \cite{y3-generalmethods}.
However, when computing the angular clustering power spectrum (Eq.~\ref{eq:gg}), the Limber approximation is insufficient , and we follow \cite{Fang_nonlimber}. For example, the exact expression for the galaxy angular power spectrum (ignoring magnification and RSD) is
\begin{align}
  \nonumber   C_{gg}^{ij} (\ell)=&\frac{2}{\pi}\int d \chi_1\,W^i_g(\chi_1)\int d\chi_2\,W^j_g(\chi_2)\\
     &\int\frac{dk}{k}k^3 P_{gg}(k,\chi_1,\chi_2)j_\ell(k\chi_1)j_\ell(k\chi_2)\,,
 \label{eq:Cl_DD}
\end{align}
and the full expressions including magnification and RSD are given in \cite{Fang_nonlimber}. Schematically, the integrand in Eq.~\ref{eq:Cl_DD} is split into the contribution from non-linear evolution, for which un-equal time contributions are negligible so that the Limber approximation is sufficient, and the linear-evolution power spectrum, for which time evolution factorizes. 

We relate the power spectra to the angular correction functions via (e.g.  \citealt{stebbins96,kamionkowski97})
\begin{align}
    w^i(\theta) =& \sum_\ell \frac{2\ell+1}{4\pi}P_\ell(\cos\theta) C^{ii}_{\delta_{\mathrm{l,obs}}\delta_{\mathrm{l,obs}}}(\ell)~,\\
    \gamma_t^{ij}(\theta) =& \sum_\ell \frac{2\ell+1}{4\pi\ell(\ell+1)}P^2_\ell(\cos\theta) C^{ij}_{\delta_{\mathrm{l,obs}}\mathrm{E}}(\ell)~,
\end{align}
where $P_\ell$ and $P_\ell^2$ are the Legendre polynomials. 
\section{Data}
\label{sec:data}

DES collected imaging data for six years, from 2013 to 2019, using the Dark Energy Camera (DECam) \citep[DECam;][]{Flaugher:2015} mounted on the Blanco 4m telescope at the Cerro Tololo Inter-American Observatory (CTIO) in Chile. The observed sky area covers $\sim 5000 \deg^2$ in five broadband filters, $grizY$, covering near infrared and visible wavelengths. This work uses data from the the first three years (from August 2013 to February 2016), with approximately 4 overlapping exposures over the full wide-field area, reaching a limiting magnitude of $i\sim23.3$ for S/N = 10 point sources. 

The data were processed by the DES Data Management system \citep{Morganson:2018} and, after a complex reduction and vetting procedure, compiled into object catalogs, using the SExtractor \citep{bertin96} software for detection on coadded images. For ease of management when performing this detection, the sky is divided into chunks 0.7306 square degrees across, which we call \emph{tiles}. This catalog includes several photometric measurements for galaxies of which the Single Object Flux (SOF) is the most accurate available.
We calculate additional metadata in the form of quality flags, survey flags, survey property maps, object classifiers and photometric redshifts to build the \texttt{Y3 Gold} data set \citep{y3-gold}. \\

\subsection{Lens samples}
This paper uses two different samples of lens galaxies:  \redmagic, a sample of Luminous Red Galaxies selected from the \redmapper galaxy cluster calibration, and \maglim, a sample with a redshift dependent magnitude limit optimized for combinations of clustering and galaxy-galaxy lensing. 

\subsubsection{\maglim}

Our fiducial sample, \maglim, is defined with a magnitude cut in the i-band that depends linearly with photometric redshift, $i < 4 z_{\rm phot} + 18$, where $z_{\rm phot}$ is the photometric redshift estimate from DNF \citep{DNF}. This selection has been optimized in \cite{y3-2x2maglimforecast} in terms of the $w$CDM cosmological constraints from the \twobytwo\ data vector, resulting in a sample with ~3.5 times more galaxies than redMaGic and ~30\% wider redshift distributions. The sample is divided in 6 tomographic bins using the the $\mathrm{DNF\_ZMEAN\_SOF}$ quantity  with bin edges $z = [0.20, 0.40, 0.55, 0.70, 0.85, 0.95, 1.05]$. 

The \maglim sample shows variations in number density correlated with observing properties which are corrected for with weights applied to each galaxy, described in \cite{y3-galaxyclustering}.

The final \maglim selection can be summarized by the following cuts on quantities from the gold catalog:
\begin{itemize}
    \item Removed objects with FLAGS\_GOLD in 2|4|8|16|32|64
    \item Star galaxy separation with EXTENDED\_CLASS = 3
    \item SOF\_CM\_MAG\_CORRECTED\_I $< 4 \ z_{DNF\_ZMEAN\_SOF} + 18$
    \item SOF\_CM\_MAG\_CORRECTED\_I $> 17.5$
    \item $0.2 < z_{\mathrm{DNF\_ZMEAN\_SOF}} < 1.05$.
\end{itemize}
See \cite{y3-gold} for further details on these quantities. 

\subsubsection{\redmagic}

We also use the DES Year 3 \redmagic\ sample. \redmagic\ selects Luminous Red Galaxies (LRGs) using the sequence model calibrated from bright red galaxy spectra, using the redmapper calibration \citep{rykoff14,rykoff16}. The \redmagic\ sample is produced by applying a redshift-dependent threshold luminosity $L_{\rm min}$ that selects for constant co-moving density. The full \redmagic\ algorithm is described in \cite{sv-redmagic}.

We divide the Y3 \redmagic\ sample into 5 photometric redshift bins, selected on the \redmagic\ redshift point estimate ZREDMAGIC. The bin edges used are $z=0.15, 0.35, 0.50, 0.65, 0.80, 0.90$. The first three bins use a luminosity threshold of $L_{\min} > 0.5 L_{*}$ and are known as the \emph{high density} sample. The last two redshift bins use a luminosity threshold of $L_{\min} > 1.0 L_{*}$ and are known as the \emph{high luminosity} sample.

The redshift distributions are computed by stacking samples from the redshift PDF of each individual \redmagic\ galaxy, allowing for the non-Gaussianity of the PDF. From the variance of these samples we find an average individual redshift uncertainty of $\sigma_z/(1+z)=0.0126$ in the redshift range used.

In \citet{y3-galaxyclustering} it was found the \redmagic\ number density correlates with a number of observational properties of the survey. This imprints a non-cosmological bias into the galaxy clustering. To account for this we assign a weight to each galaxy which corresponds to the inverse of the angular selection function at that galaxies location. The computation and validation of these weights is described in \cite{y3-galaxyclustering}.  


The final \redmagic\ selection can be summarized by the following cuts on quantities from the gold catalog and \redmagic\ calibration,
\begin{itemize}
    \item Removed objects with FLAGS\_GOLD in 8|16|32|64
    \item Star galaxy separation with EXTENDED\_CLASS >= 2 
    \item Cut on the red-sequence goodness of fit $\chi^{2} < \chi^{2}_{\rm max}(z)$
    \item $0.15 < \mathrm{ZREDMAGIC} < 0.9$
\end{itemize}

The star galaxy separator EXTENDED\_CLASS is defined as the sum of three integer conditions, $T+5T_{\rm err} > 0.1$, $T+T_{\rm err} > 0.05$, and $T-T_{\rm err} > 0.02$, where $T$ is the galaxy size squared, as determined by the SOF composite model described in \cite{y3-gold} measured in ${\rm arcmin}^2$.

\subsection{Mask}

The lens samples are selected from within the DES Year 3 \threebytwo\ footprint, defined on a pixelated healpix map \citep{healpix} with $N_{\rm side}=4096$. This angular mask only includes pixels with photometry deep enough  that both lens samples are expected to have a uniform selection function in all redshift bins. We also remove pixels close to foreground objects, with photometric anomalies, or with with a fractional coverage less than 80\%, resulting in a total area of $4143\mathrm{deg}^2$. The GOLD catalog quantities we select on are summarized by, 
\begin{itemize}
    \item footprint >= 1
    \item foreground == 0
    \item badregions <= 1
    \item fracdet > 0.8
    \item depth\_i >= 22.2
    \item ZMAX\_highdens > 0.65
    \item ZMAX\_highlum > 0.95 
\end{itemize}

See \cite{y3-gold} for further details on these quantities.

\subsection{Source sample}

The source sample is another subset of the DES Year 3 Gold catalog~\citep{y3-gold}. It consists of 100,208,944 galaxies with measured photometry and shapes after imposing the following cuts in $r, i,$ and $z$ bands, as motivated in \citet*{y3-shapecatalog}:
\begin{itemize}
\item $18 < m_i<23.5$
\item $15<m_r<26$
\item $15 < m_z<26$
\item $-1.5 < m_r-m_i < 4$
\item $-4 < m_z-m_i < 1.5$
\end{itemize}
The shapes of these galaxies, determined in \cite*{y3-shapecatalog} and calibrated for use in weak lensing shear statistics in \cite{y3-imagesims}, are used for the galaxy-galaxy lensing measurement. This measurement also requires the redshift distribution of the source galaxies. Just as the lens galaxies are divided into distinct redshifts bins, the source galaxies are divided into 4 redshift bins, with mean redshifts ranging from 0.34 to 0.96. \citet*{y3-sompz} describes how these bins are populated and the inference of the redshift distributions and uncertainties for each bin.

\section{Simulations}
\label{sec:simulations}

A number of simulations are used in this analysis. The details of these simulations are described here.

\subsection{\balrog}
\label{sec:sims-balrog}


The \balrog image simulations are created by injecting `fake' galaxy images into real DES single-epoch wide-field images. The complete DES photometric pipeline is run on the images, resulting in object catalogs.
The objects in the output catalogs can be matched to the \balrog injections to investigate the survey transfer function. The injected galaxies are model fits to the DES deep-field observations which are typically 3-4 magnitudes deeper than the wide field data \citep*{y3-deepfields}. Further details of the Year 3 \balrog simulations are described in \cite{y3-balrog}. 

A number of \balrog catalogs were produced for the DES Year 3 analysis. In this analysis we use \balrog run2a and run2a-mag. These runs both cover the same 500 random DES tiles, with approximately 4 million detected objects in each. The injections were randomly selected from objects in the DES deep fields down to a magnitude limit of 24.5.\footnote{This was in fact a magnitude based on the total flux in the $riz$-bands - see \citet{y3-balrog} for details.} In run2a-mag, the exact same deep field objects are injected at the same coordinates as in run2a but with a 2\% magnification applied to each galaxy image.\footnote{Magnification is applied to the injected images using the GalSim \citep{galsim} \texttt{magnify} method.} This magnification increases the size of the image by 2\% while preserving surface brightness such that, in the absence of systematics and selection effects, the flux is also expected to increase by 2\%.

\begin{figure}
    \centering
    \includegraphics[width=\linewidth]{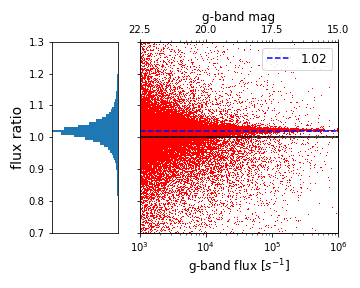}
    \caption{ Ratio of \balrog fluxes, measured in counts per second, in the magnified and unmagnified \balrog runs. The average flux difference is consistent with the input 2\% magnification. The large scatter at low flux  is dominated by noise in the SOF photometric fitting. the left panel is a histogram of the flux ratio showing the tails are small and the distribution is centered on 1.02. }
    \label{fig:balrog_mag_flux}
\end{figure}

All Sextractor and SOF quantities used in the lens sample selection are computed on the matched objects in both run2a and run2a-mag. The difference in $g-$band fluxes for the same objects is shown in \fig{fig:balrog_mag_flux}. The scatter in the flux difference is dominated by noise in the photometric fitting. 

We show in Appendix~\ref{app:balrog_plots} that the \balrog\ method produces a realistic simulation of the DES-Y3 data,  with good agreement in distributions of measured quantities such as magnitudes, sizes and photometric redshifts. 

\subsection{N-body simulations}\label{sec:nbody_description}
\subsubsection{\textsc{Buzzard v2.0}}\label{sec:buzzard_description}
The \textsc{Buzzard v2.0} simulations are a suite of 18 synthetic DES Y3 galaxy catalogs constructed from $N$-body lightcone simulations~\citep{y3-simvalidation}. Each pair of two synthetic DES Y3 catalogs is generated from a set of three independent lightcones with mass resolutions $[0.33,\, 1.6, \, 5.9] \, \times10^{11}\, h^{-1}M_{\odot}$, box sizes of $[1.05,\, 2.6,\, 4.0]\, (h^{-3}\, \rm Gpc^3)$, spanning redshift ranges in the intervals $[0.0,\, 0.32,\, 0.84, \,  2.35]$ respectively. Each lightcone is run with \textsc{L-Gadget2}, a version of \textsc{Gadget2} \citep{Springel2005} optimized for memory efficiency when running dark-matter-only configurations. The simulations were initialized at $z=50$ with initial conditions generated by \textsc{2LPTIC} \citep{Crocce2005} from linear matter power spectra produced by \textsc{CAMB} \citep{Lewis2000} at the \buzzard cosmology. 

Galaxies are added to the $N$-body outputs using the \textsc{Addgals} algorithm \citep{Wechsler2021,DeRose2021a}, which imbues each galaxy with a position, velocity, absolute magnitude, spectral energy distribution, half-light radius and ellipticity. The \textsc{Calclens} raytracing code \citep{Becker2013}, which employs a spherical harmonic transform Poisson solver on an $N_{\rm side}=8192$ \textsc{HEALPix} grid \citep{healpix}, is used to compute lensing quantities, including convergence and shear, at each galaxy position. These quantities are then used to magnify galaxy magnitudes and sizes, and shear ellipticities. The catalogs are cut to the DES Y3 footprint and photometric errors are applied to the magnitudes using error distributions derived from \textsc{Balrog} \citep{y3-balrog}. 

The \redmagic sample is selected from each synthetic galaxy catalog using the same algorithm that is employed on the DES Y3 data. This is possible given the close match between red-sequence galaxy colors in \buzzard\ and the DES Y3 data. A source galaxy sample is selected to match the effective number density and shape noise of the DES Y3 \textsc{Metacalibration} source sample \citep*{y3-shapecatalog}, and photometric redshifts are estimated with the \textsc{SOMPZ} algorithm \citep*{y3-sompz}. For a comprehensive overview of these simulations, see \citet{y3-simvalidation}.

\subsubsection{MICE}\label{sec:mice_description}

The MICE Grand Challenge (MICE-GC) simulation is a large N-body run which evolved $4096^3$ particles in a volume of $(3072 \, {\rm Mpc} \, h^{-1})^3$ using the \textsc{gadget-2} code \citep{Springel2005} \citep{2015MNRAS.448.2987F}. It assumes a flat $\Lambda$CDM cosmology with $\Omega_m=0.25$, $\Omega_{\Lambda}=0.75$, $\Omega_b=0.044$, $n_s=0.95$, $\sigma_8=0.8$ and $h=0.7$. This results in a particle mass of $2.93 \times 10^{10} \, h^{-1}M_{\odot}$ \citep{2015MNRAS.448.2987F}.  The run produced on-the-fly a light-cone output of dark-matter particles up to $z=1.4$ without repetition in one octant \citep{2015MNRAS.453.1513C}. A set of 256 maps of the projected mass density field in narrow redshift shells, with angular Healpix resolution $N_{side}=8184$, were measured. These were used to derive the convergence field $\kappa$ in the Born approximation by integrating them along the line-of-sight weighted by the appropriate lensing kernel. These $\kappa$ maps are then used to implement the magnification in the magnitudes and positions of mock galaxies due to weak lensing, as detailed in \cite{2015MNRAS.447.1319F}.


Haloes in the light-cone were 
populated with galaxies as detailed in \cite{2015MNRAS.447..646C}, assigning positions, velocities, luminosities and colours to reproduce the luminosity function, (g-r) color distribution and clustering as a function of color and luminosity in SDSS \citep{2003ApJ...592..819B,2011ApJ...736...59Z}. Spectral energy distributions (SEDs) are then assigned to the galaxies resampling from the COSMOS catalogue of \cite{2009ApJ...690.1236I} galaxies with compatible luminosity and (g-r) color at the given redshift. Once with the SED any desired magnitude can be computed. In particular, DES magnitudes are generated by convolving the SEDs with the DES pass bands including the expected photometric noise per band given the depth of DES Y3. Finally, in order to reproduce with high fidelity the distribution of colours and magnitudes of the observational data with map data photometry into the MICE one using an N-dimensional PDF transfer function, which also preserves the correlation among colours. 

Once provided with this catalogue we run both the Redmagic and DNF algorithms to determine photometric redshifts, starting from magnitudes with and without the contribution from magnification. We then selected the \redmagic and \maglim samples. The abundance, clustering and photometric redshift errors of the real and simulated data resemble each other very well for both samples. 
\section{Estimating magnification coefficients}
\label{sec:method}


As described in \sect{sec:magnification_theory}, the constant $C$ in Eq.~(\ref{eq:definition_C}), which is the response of the galaxy number density to $\kappa$, can be split into $C=\csample+\carea$. The $\carea$ from the area change is equal to $-2$. The $\csample$ from the flux and galaxy size change can be estimated from our simulations separately, as the fractional change in the number of selected galaxies in response to a small convergence, $\delta\kappa$, applied to the simulated, input galaxy properties (i.e.\ flux and size) only --- note the galaxy positions are not altered and so the change in area effect is not included. $\csample$ can then be estimated simply via a numerical derivative 
\begin{equation}
    \csample \delta\kappa = \frac{N(\delta\kappa)-N(0)}{N(0)},
\end{equation}
where $N(0)$ and $N(\delta \kappa)$ are the 
absolute number of galaxies selected from the $\kappa=0$ and $\kappa=\delta\kappa$ simulations respectively. Then
\begin{equation}
     \csample = \frac{N(\delta\kappa)-N(0)}{N(0)\delta\kappa}.
     \label{eq:c_estimate}
\end{equation}
This is the basic equation we will use to estimate $\csample$, but using a variety of input data, as described in Secs.~\ref{sec:balrog}-\ref{sec:magsims}.

\subsection{ Estimate from \Balrog simulations}\label{sec:balrog}

The \balrog magnification run described in section \ref{sec:sims-balrog} uses the same input galaxy models in the same positions as the unmagnified run, but with a constant magnification $\delta \mu=1.02$ (i.e. $\delta \kappa \sim 0.01$) applied to each input galaxy. We find $\delta\kappa \sim 0.01$ is large enough that we can get a sufficiently precise estimate of $\csample$ (i.e.\ a sufficient number of objects are magnified across the detection threshold), but small enough to ensure that the quadratic $~\kappa^2$ contributions to the change in number density are small ($\sim 10^{-4}$).
We apply the \maglim\ and \redmagic\ lens sample selection on the galaxy catalogs from both the $\kappa=0$  run  and the $\kappa=\delta\kappa$ run. We then estimate $\csample$ via Eqn.~(\ref{eq:c_estimate}).


This estimate should capture the impact of magnification on the specific colour and magnitude selection of the \redmagic and \maglim samples, plus any size selections such as the star-galaxy separation cuts.

\begin{figure*}
    \centering
    \includegraphics[width=\textwidth]{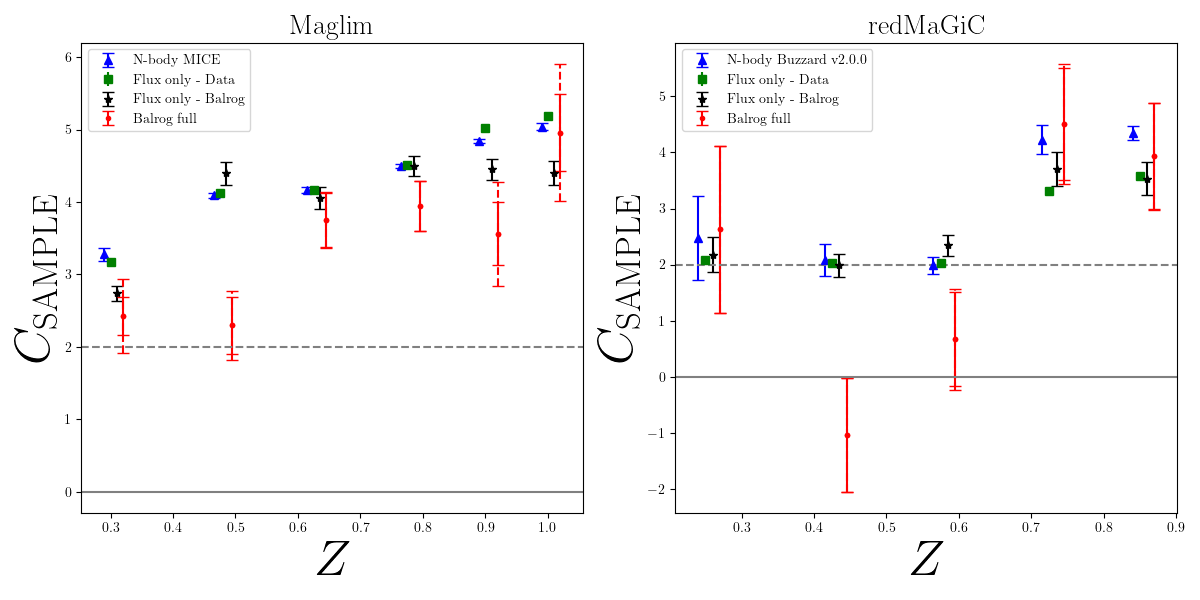}
    \caption[width=\textwidth]{Magnification coefficient estimates for the two lens samples. Each panel shows multiple estimates for the magnification coefficients from the different methods outlined in \sect{sec:method}. Our primary method of estimating these coefficients, shown as red circles, uses the \balrog\ simulations (with and without a small magnification applied to the injected galaxy properties), to accurately quantify galaxy selection effects and systematic effects (as described in \sect{sec:balrog}). The blue triangles show an estimate from N-body simulations, containing flux magnification only (see \sect{sec:magsims}). The green squares are estimates from  perturbing the measured fluxes in the data (see \sect{sec:data_method}). The black stars are from  perturbing the measured fluxes in the baseline simulated \Balrog sample. If the \balrog sample was truly representative of the real data we would expect the green and black points to be the same. We therefore use the difference between the green and black points is used as a source of systematic error on the red \balrog estimates. We show both the statistical errors and the total (stat + sys) error from \balrog. The solid line corresponds to zero magnification bias from the sample selection, while the dashed line corresponds to zero magnification bias when also including the change in area element. }
    \label{fig:c}
\end{figure*}

The estimates of $\csample$ in each of the tomographic bins for the two lens samples using \balrog\ are shown in \fig{fig:c}, labelled `Balrog full'. These estimates are subject to shot-noise due to the finite volume of the Balrog simulation, which we calculate as:

\begin{equation}
    \label{eq:c_err_balrog}
    \frac{\sigma^{\rm stat}}{\cbalrog} = 
    \sqrt{\frac{ N(0 \text{ only}) + N(\delta\kappa \text{ only})}{\left[N(\delta\kappa) - N(0)\right]^2} + \frac{1}{N(0)} + \frac{2N(0 \text{ only})} {N(0)\left[N(\delta\kappa) - N(0)\right]}}.
\end{equation}
where $N(0 \text{ only})$ is the number of objects selected from the $\kappa=0$ simulation and not selected from the $\kappa=\delta\kappa$ simulation, and $N(\delta\kappa \text{ only})$ is the number of objects selected from the $\kappa=\delta\kappa$ simulation and not selected from the $\kappa=0$ simulation.
This is the statistical contribution to the error bars shown in \fig{fig:c}. A derivation of this uncertainty can be found in Appendix \sect{app:stat_error}.

\subsection{ Estimate from perturbing  measured fluxes }
\label{sec:data_method}

For this method we add a constant offset $\Delta m$ directly to the Y3 real data magnitudes used in the sample selection.
\begin{equation}
    \label{eq:delta_magnitude}
    \Delta m = -2.5 \log_{10}( 1 + 2 \delta \kappa )
\end{equation}
where $\delta \kappa=0.01$.

We then re-select the sample using these perturbed magnitude and can compute the $\cdata$ value directly from Eqn. \ref{eq:c_estimate}. Note that we do not re-run the photometric redshifts so these are computed using quantities derived from the true magnitudes. 
This method provides a simplistic estimate of the effect of magnification on the fluxes only and ignores the effects of photometric noise, selection on photometric redshift, size selection, observational systematics, and more generally the survey transfer function. 

Since there is no additional photometric noise introduced between the original and perturbed fluxes in this method, we can use a simplified version of Eq.~\ref{eq:c_err_balrog} to estimate the statistical uncertainty, 
\begin{equation}
    \label{eq:c_err}
    \frac{\sigma_{\cdata}}{\cdata} = 
    \sqrt{ \frac{1}{N(\delta\kappa) - N(0)} + \frac{1}{N(0)}}.
\end{equation}

We also apply this `flux-only' method to the \Balrog catalogs; by comparing to the `flux-only' results on the real data, this tests how representative the \Balrog sample is of the real data.

\subsection{N-body simulations}\label{sec:magsims}

The third method used in this analysis is performed on the MICE and Buzzard N-body simulations described in Sec.~\ref{sec:buzzard_description} and Sec.~\ref{sec:mice_description}. This method takes advantage of of the fact that we know the true convergence, $\kappa$, at each simulated galaxy location. The N-body simulations used here include magnification effects on the galaxy positions and fluxes, but do not realistically simulate the full impact of lensing on observed galaxy images, and our estimate of the selection response from them includes only that due to change in flux. 

We select the lens samples with and without magnification applied to the fluxes and compute fractional change in number of selected objects, in 10 equally spaced $\kappa$ bins. From Eq.~\ref{eq:c_estimate} one can see that the gradient of this relation is equal to $\csample$. We use estimate this gradient with a least square fit and use this as $\cnbody$. 

As with our estimate from perturbing measured fluxes in the data \sect{sec:data_method}, this method only captures the effect of magnification on galaxy fluxes, and not any size selection effects. 
We also find  agreement with the color-magnitude distribution in the N-body simulations and the data and therefore this would also be a reasonable estimate of the data coefficients if other selection effects are small. 

We note that we initially used only the Buzzard simulations for both the \maglim\ and \redmagic\ samples, but based on evidence that the \maglim\ sample was not sufficiently representative of the real data (due to a decrement of galaxies at high redshift in Buzzard, see \citealt{DeRose2021a}), we instead used the MICE simulations for \maglim\ only.

\subsection{Comparison of magnification coefficient estimates}
\label{sec:cflux_discussion}

Tables \ref{table:maglim} and \ref{table:redmagic} show the relevant number counts and $\csample$ estimates from the three different methods. The $\csample$ estimates are also compared in \fig{fig:c}. The $\cbalrog$ estimates include a systematic error that accounts for any differences in the color-magnitude-size selection in \balrog compared to the real data. We compute this by running the flux-only estimate described in Sec.~\ref{sec:data_method} on the unmagnified \balrog sample, and take the difference between this estimate and the flux-only estimate from the data to be a systematic error in the $\cbalrog$ estimates. This difference captures only differences in the flux-size distribution, not the effects from our magnified \balrog method. In general the distribution of \balrog magnitudes and sizes agrees well with the data, as can be seen in Appendix \ref{app:balrog_plots}, but there are some small differences at the size selection cut. The flux-only \csample estimates on \balrog are shown in Fig,~\ref{fig:c} and agree well with the flux-only data estimates for the \redmagic sample. The systematic error is largest for the \maglim sample in bins 1, 5 and 6, potentially indicating that the \balrog \maglim sample is less representative of the real data at these redshifts. 


The \csample estimates from the magnified \balrog sample (our fiducial measurement) tend to be smaller than the flux-only methods. This is particularly apparent in redshift bins between ${\sim}0.3$ and ${\sim}0.6$. This difference could be caused by the dependence on quantities other than flux inherent in the sample selection, for example size (used in star-galaxy separation), or systematics in the photometric pipeline. We note that the agreement is better for simple flux limited samples without tomographic binning,  as shown in \citet{y3-balrog}. 

The \redmagic\ coefficients tend to have smaller magnification bias than \maglim, and when including the $C_{\rm area}$ contribution, the low-redshift \redmagic\ total magnification bias contributions are small. 

We believe \balrog, which include a wide range of observational effects, and account for selection on quantities beyond only flux, is our most accurate method for estimating the magnification coefficients, so use the \balrog estimates as fiducial 
hereafter. 

\begin{table*}
    \centering
    \begin{tabular}{|l|l|l|l|}
\hline 
    \multicolumn{4}{|c|}{\textbf{\maglim}}   \\ \hline 
    redshift & \cbalrog & \cdata & \cnbody \\ \hline
0.2 < z < 0.4   & $2.43\pm0.26 \pm 0.44 \text{ (sys)}$ & $3.18  \pm0.012$   &$3.28 \pm0.091$  \\
0.4 < z < 0.55  & $2.30\pm0.39 \pm 0.27 \text{ (sys)}$ & $4.13  \pm0.016$   &$4.09 \pm0.034$  \\
0.55 < z < 0.7  & $3.75\pm0.38 \pm 0.11 \text{ (sys)}$ & $4.17  \pm0.016$   &$4.17 \pm0.04$  \\
0.7 < z < 0.85  & $3.94\pm0.35 \pm 0.01 \text{ (sys)}$ & $4.52  \pm0.015$   &$4.5  \pm0.03$  \\
0.85 < z < 0.95 & $3.56\pm0.44 \pm 0.57 \text{ (sys)}$ & $5.02  \pm0.018$   &$4.84 \pm0.023$  \\
0.95 < z < 1.05 & $4.96\pm0.53 \pm 0.79 \text{ (sys)}$ & $5.19  \pm0.019$   &$5.04 \pm0.054$  \\
\hline
    \end{tabular}
    \caption{$\csample$ estimates for the \maglim\ sample, for the three different methods described in \sect{sec:method}. The \balrog\ estimate include a systematic uncertainty derived the difference between the perturbation of measured fluxes method applied to the data and \balrog\ samples. 
    }
    \label{table:maglim}
\end{table*}

\begin{table*}
    \centering
    \begin{tabular}{|l|l|l|l|}
\hline 
    \multicolumn{4}{|c|}{\textbf{\redmagic}}   \\ \hline 
    redshift & \cbalrog & \cdata & \cnbody \\ \hline
0.15 < z < 0.35 &  $2.63 \pm1.50 \pm 0.093 \text{ (sys)}$  & $2.08  \pm0.025$   &$2.47 \pm0.753$  \\
0.35 < z < 0.5  & $-1.04 \pm1.01 \pm 0.32  \text{ (sys)}$  & $2.02  \pm0.019$   &$2.08 \pm0.287$  \\
0.5 < z < 0.65  &  $0.67 \pm0.90 \pm 0.32  \text{ (sys)}$  & $2.03  \pm0.015$   &$1.99 \pm0.157$  \\
0.65 < z < 0.8  &  $4.50 \pm1.07 \pm 0.39  \text{ (sys)}$  & $3.32  \pm0.027$   &$4.22 \pm0.259$  \\
0.8 < z < 0.9   &  $3.93 \pm0.95 \pm 0.043 \text{ (sys)}$  & $3.58  \pm0.031$   &$4.35 \pm0.122$  \\
\hline
    \end{tabular}
    \caption{$\csample$ estimates for the \redmagic\ sample, for the three different methods described in \sect{sec:method}. The \balrog\ estimate include a systematic uncertainty derived the difference between the perturbation of measured fluxes method applied to the data and \balrog\ samples. 
    }
    \label{table:redmagic}
\end{table*}

\section{Expected impact on DES Y3 analyses}
\label{sec:simresults}

In this section we estimate the impact of magnification in the DES Year 3 galaxy clustering + galaxy-galaxy lensing (\twobytwo) analysis, using a noiseless datavector generated from our theoretical model (we use the same fiducial model and parameter values as \citealt{y3-generalmethods}). For these tests we use the default $\csample = \cbalrog$ values estimated with the \balrog simulations. 

\begin{figure*}
    \centering
    \subfigure{\includegraphics[width=\linewidth]{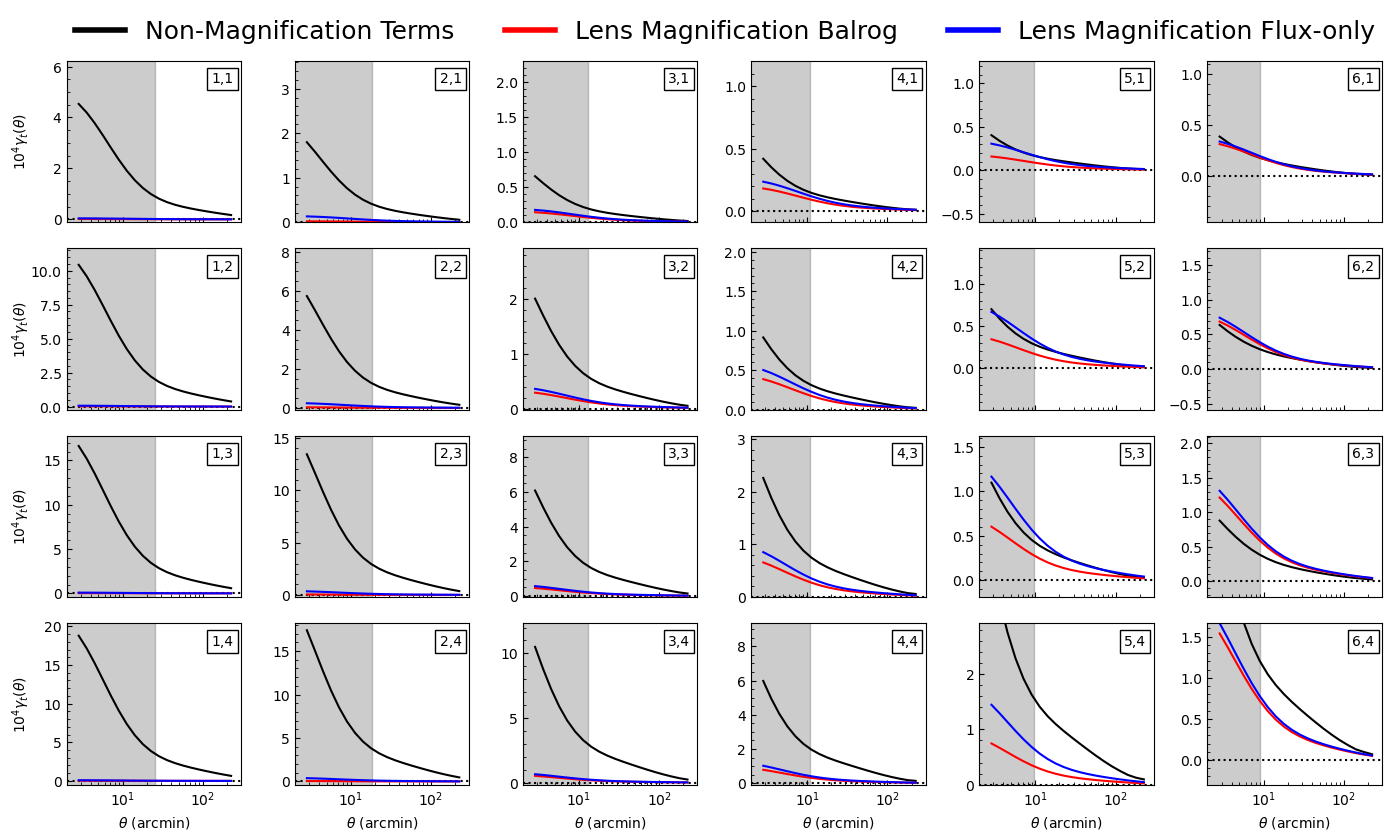}}\quad
    \subfigure{\includegraphics[width=\linewidth]{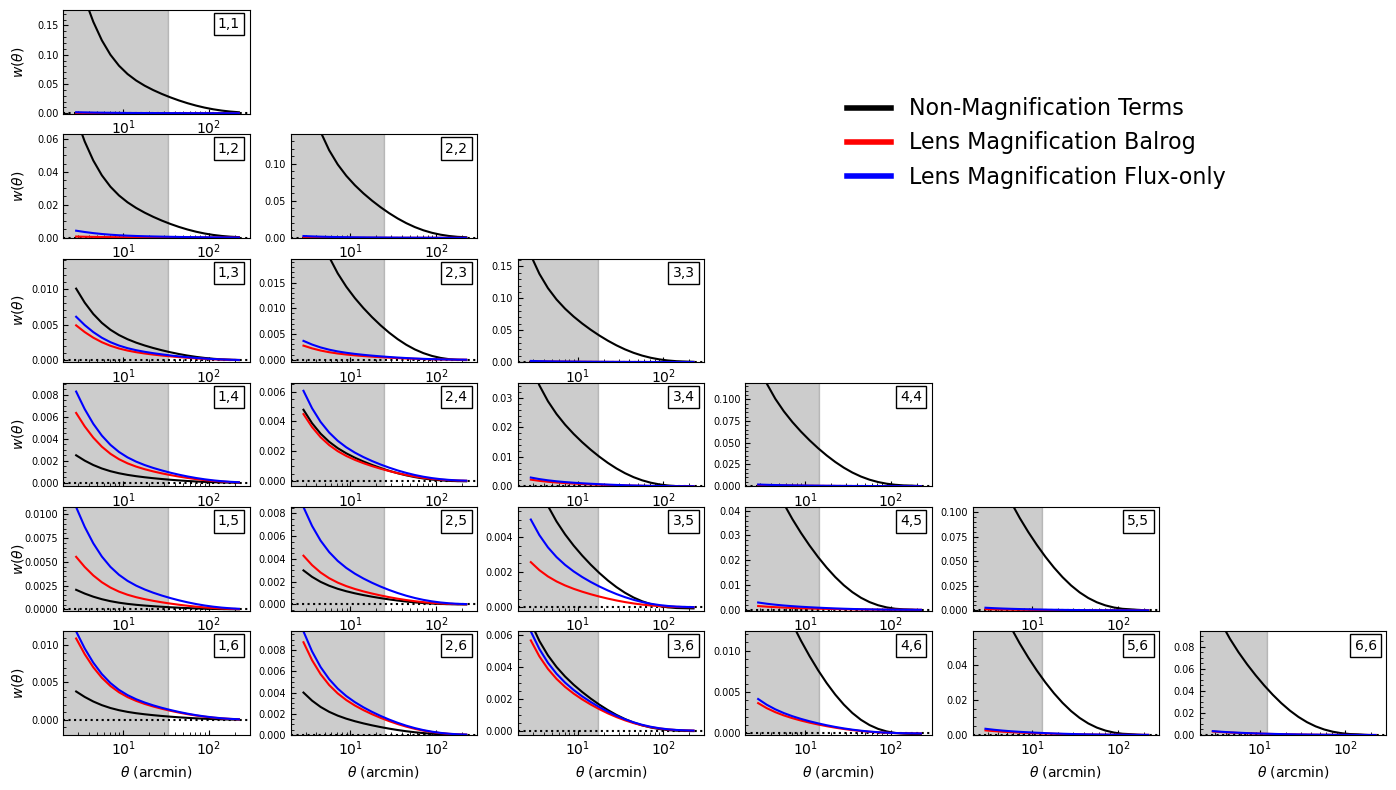}}

    \caption{The impact of magnification on the galaxy-galaxy lensing (\gammat) (top panel) and galaxy clustering (\wtheta) (bottom panel) model prediction, for all redshift bin combinations, for the \maglim\ sample. In each panel, the black line (labelled `Non-Magnification Terms') is the model prediction when ignoring lens magnification terms. The red and blue lines are the additional contributions to the model prediction from lens magnification, for magnification coefficient values estimated using \balrog\ (red lines, see \sect{sec:balrog} for details), and the perturbing flux method (blue lines, see \sect{sec:data_method} for details). }
    \label{fig:theory_dvec}
\end{figure*}

To guide intuition for the subsequent analysis of parameter biases, we show in Fig.~\ref{fig:theory_dvec}  the impact of lens magnification on the different parts of the DES Year 3 data vector (for the \maglim\ sample). It shows magnification has the largest impact on the galaxy-galaxy lensing of high redshift source bins around high redshift lens bins. This is expected since only high redshift lens galaxies will experience large magnification.
In relative terms, the clustering auto-correlations have a small contribution due to magnification, while for widely separated redshift bins (e.g.\ the (1,6) pairing), magnification is the dominant contribution to the signal. Despite this impact, we note that we might still expect little impact from lens magnification on the cosmological parameter constraints for the fiducial DES Y3 cosmology analysis in \cite{y3-3x2ptkp} because i) most of signal-to-noise in the galaxy-galaxy lensing datavector is contributed by the lowest three lens redshift bins where biases are small 
 and ii) the cross-correlation clustering signal between different lens redshift bins is not used in the fiducial \threebytwo\ analysis in \cite{y3-3x2ptkp} (though we note \cite{thiele_2020} find magnification bias can still be significant in the absence of cross-redshift bin clustering due to the cosmological bias from the cross-correlations acting in the opposite direction to the galaxy-galaxy lensing).  We do consider this cross-correlation signal in this work, given its constraining power on the magnification signal.

Having generated a noiseless datavector from our fiducial theoretical model, we perform cosmological parameter inference 
following the DES Year 3 analysis choices outlined in \cite{y3-generalmethods}. We test this within a \lcdm\ model. This analysis is designed to estimate the impact of magnification analysis choices in an idealised case, where we can identify the expected size of the magnification signal and the expected projection effects from new parameters. We later explore magnification choices on N-body simulations and the real data. 

\begin{figure*}
    \centering
    \subfigure{\includegraphics[width=0.4\linewidth]{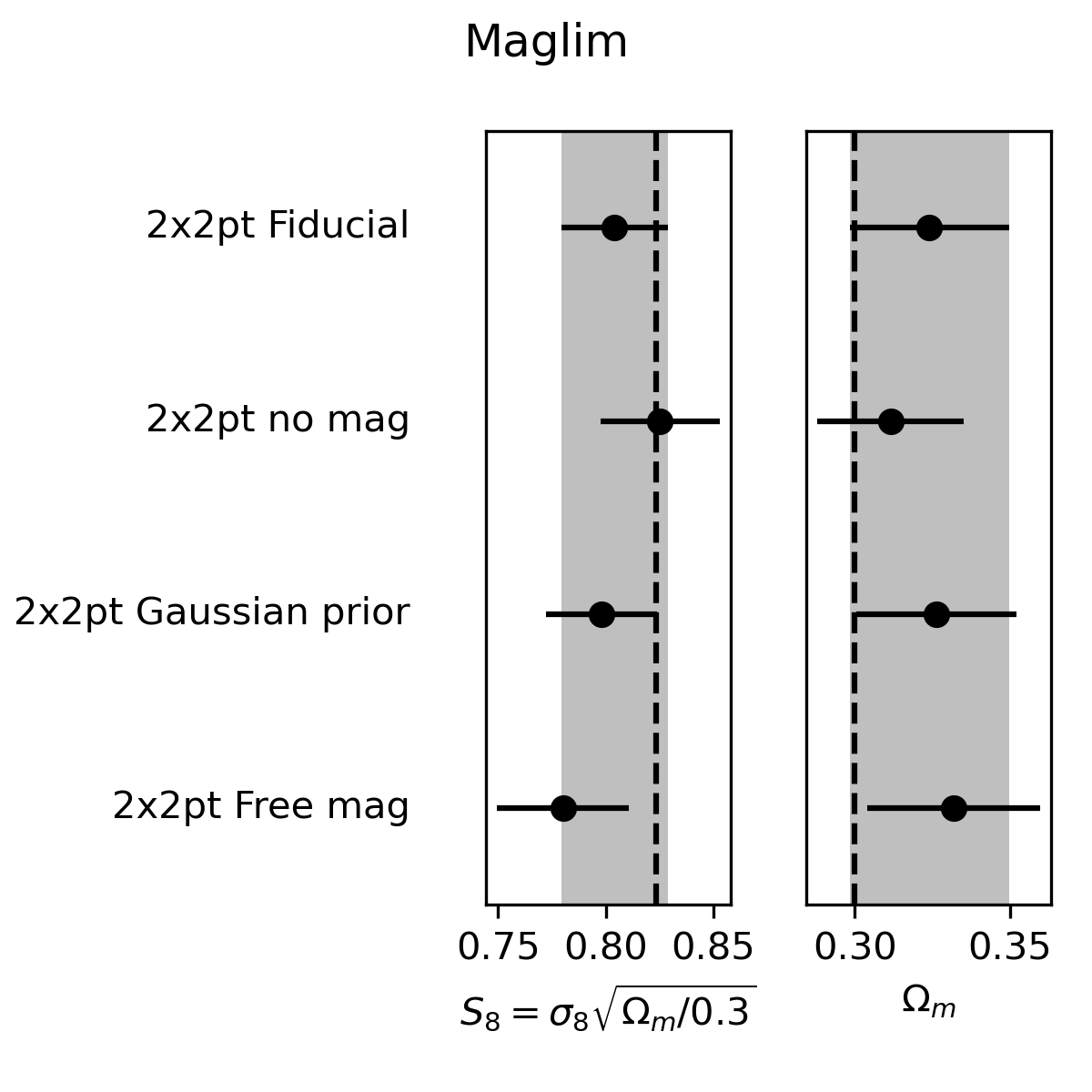}}\quad
    \subfigure{\includegraphics[width=0.4\linewidth]{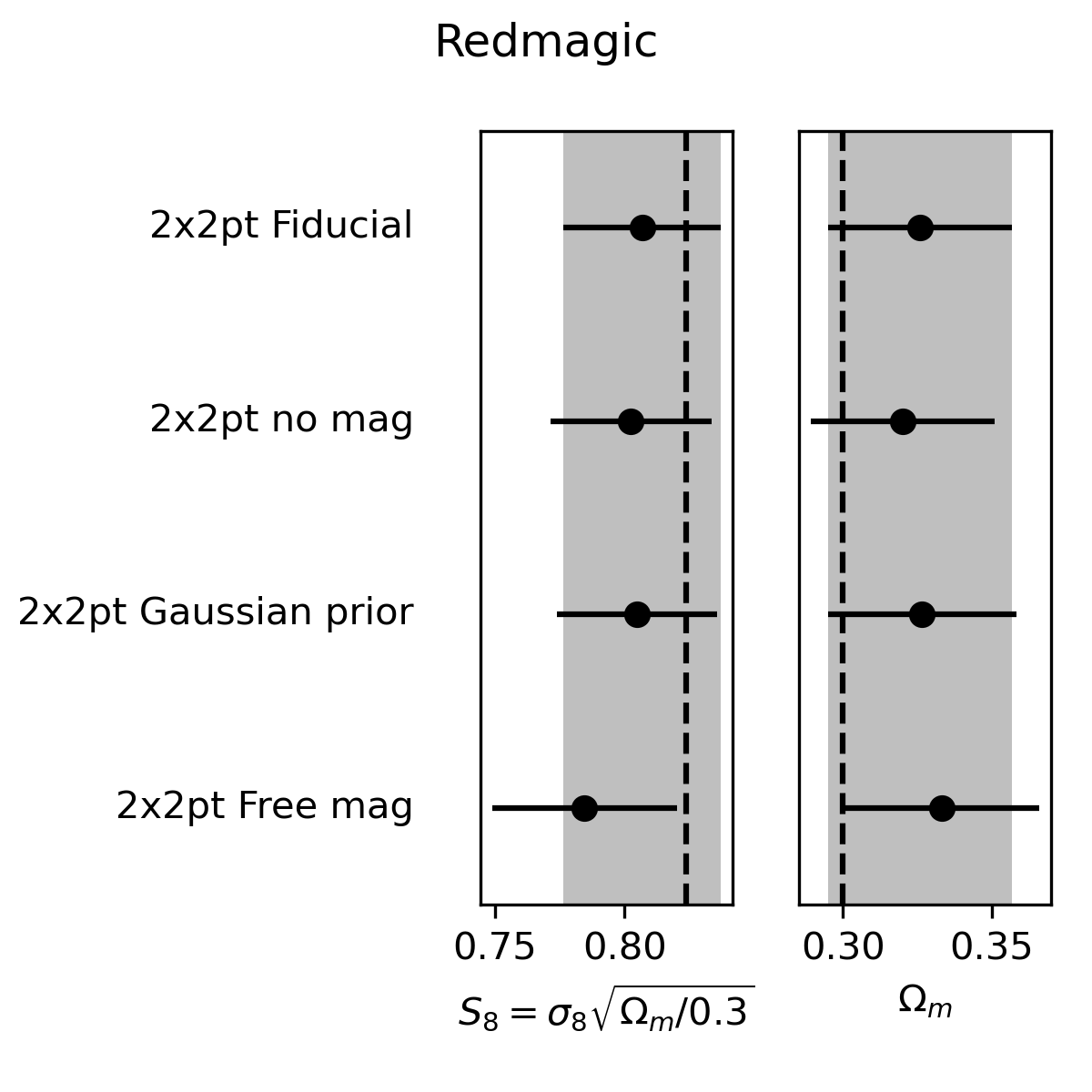}}
    \caption{ \lcdm\ parameter constraints from a simulated, noiseless, galaxy clustering + galaxy-galaxy lensing data vector closely following the DES Year 3 methodology with both the \maglim lens sample (left) and the  \redmagic lens sample (right). Three different priors on the magnification bias $\csample$ are shown; a delta function at the true values used in the simulated vector (labelled ``Fiducial"), a flat prior with width (green), a Gaussian prior with width equal to the \balrog errors in Table \ref{table:redmagic}, and wide, uniform priors $-4<\csample<12$ (labelled `Free mag'). The input magnification coefficients are the \balrog values. The dashed vertical line indicates true input values to the datavector, and shaded band the $1\sigma$ uncertainty region for the Fiducial analysis.}
    \label{fig:sim_c_priors}
\end{figure*}

We show in \fig{fig:sim_c_priors} the recovered constraints on $S_8=\sigma_8\sqrt{\Omega_m/0.3}$ with four choices of prior on the magnfication coefficients $\csample^i$:
\begin{enumerate}
\item{$\csample^i$ fixed to their true (i.e. used to generate the datavector) values (labelled ``\twobytwo\ Fiducial'').}
\item{$\csample^i=2$, such that there is no magnification contribution to the datavector (since $\ctotal=0$, labelled `no mag').}
\item{Gaussian priors on $\csample^i$ centered on their true values and with widths equal to the statistical uncertainties in the \balrog measurements,  listed in Tables \ref{table:redmagic} and \ref{table:maglim} (labelled ``\twobytwo\ Gaussian prior'').}
\item{Uniform priors on the $\csample^i$ in the range -4 to 12 (labelled ``\twobytwo\ Free mag'').}
\end{enumerate}

We firstly note that even for a noiseless datavector generated from the theoretical model, the mean of the marginalized posterior does not perfectly recover the input parameter values. For the \redmagic sample with fixed magnification coefficients, the mean of the $S_8$ posterior is biased (with respect to the input value)  by $-0.55\sigma$ and the mean of the $\Omega_m$ posterior is biased by $0.84\sigma$. For \maglim, the corresponding biases are $-0.79$ and $0.93\sigma$. We have verified that the maximum posterior parameter values match the input cosmological parameters to high precision, implying that the biases seen in the first row of \fig{fig:sim_c_priors} are due to `prior volume' of `projection effects'. Put broadly, these can occur when the data is not powerful enough to make the prior choice on marginalized parameters irrelevant, and their presence here mean marginalized parameter constraints should be interpreted with caution (especially when comparing to other cosmological datasets). In this case, the bias is at least partially caused by the degeneracy between $S_8$ and the sum of the neutrino masses $\sum m_{\nu}$, for which the input value $\sum m_{\nu}=0.06~{\rm eV}$ is at the lower edges of a flat prior. 

Beyond the biases from prior volume effects, there is an small \emph{additional} $-0.13\sigma$ bias in $S_8$ for the case where magnification is not included in the datavector for the \redmagic sample, while for \maglim the additional bias is larger, at  $0.85\sigma$ with respect to the case where the correct magnification coefficients are assumed in the datavector. This implies that especially for the \maglim sample, it is important to include magnification contributions in the theoretical model for our analysis.

When marginalizing over the magnification coefficients with Gaussian priors, the recovered constraints are very similar to the case where they are fixed to their true values (with an $0.06\sigma$ and $0.19\sigma$ change in $S_8$ for \redmagic and \maglim respectively, compared to the fixed magnification coefficients case). 

We find using a flat prior on $\csample$ somewhat degrades the constraining power of the analysis (the $1\sigma$ uncertainty in $S_8$ is $18\%$ and $24\%$ larger for \redmagic\ and \maglim respectively), and there is a corresponding increase in the prior volume bias in the marginalized $S_8$, which in this case is $-1.1\sigma$ ($-1.4\sigma$) for \redmagic (\maglim). 

We consider the \balrog estimates with Gaussian priors to be the most complete way to model magnification as these consider the widest range of magnification effects and their uncertainties, as shown in Sec.~\ref{sec:method}. We consider the flat prior to be the case most insensitive to the measurement of the coefficients, and the fixed case to be most convenient for running the inference pipeline. Because the flat prior induces additional prior volume effects, and the difference between Gaussian and fixed priors is negligible, these results to justify the decision to keep the magnification coefficients fixed in the fiducial DES Y3 3$\times$2pt cosmology analysis \citep{y3-3x2ptkp}. 

These results show the impact of magnification in DES Year 3 is a significant systematic that must be accounted for in the modelling of the two-point functions, in order to avoid cosmological bias. This supports similar conclusions in recent magnification studies for other survey specifications \citep{Duncan_2022,mahoney22}. 



\section{Validation using Buzzard simulations}
\label{sec:buzzard_validation}

In this section we validate our theoretical predictions and parameter inference using the DES Year 3 Buzzard simulations, which are described in detail in  \sect{sec:buzzard_description}. 
We measure a DES Y3-like datavector, consisting of tomographic galaxy clustering (\wtheta), galaxy-galaxy lensing (\gammat), and shear (\xipm) two-point correlation functions. We test our modeling of these signals in the Buzzard simulations by analysing the mean datavector across all realizations of the Y3 Buzzard simulations, while assuming a covariance on that datavector appropriate for one realization i.e. a level of uncertainty close to that for DES Y3. While the baseline DES Y3 3$\times$2pt analysis is validated in \citet{y3-simvalidation}, here we aim to test the extended analysis explored here, as well as provide further confidence that the magnification components of the signal in the baseline analysis are being modeled accurately. In particular, we test:
\begin{enumerate}
    \item{whether at fixed cosmology, the values of the magnification coefficients estimated in \sect{sec:magsims} are recovered accurately}
    \item{whether the true Y3 Buzzard cosmology is recovered accurately for different choices of prior on the magnification coefficients.}
    \item{whether the true cosmology is recovered accurately, and with what extra precision, when including cross-correlations between different redshift bins in \wtheta.}
\end{enumerate}

\begin{figure*}
\includegraphics[width=0.4\linewidth]{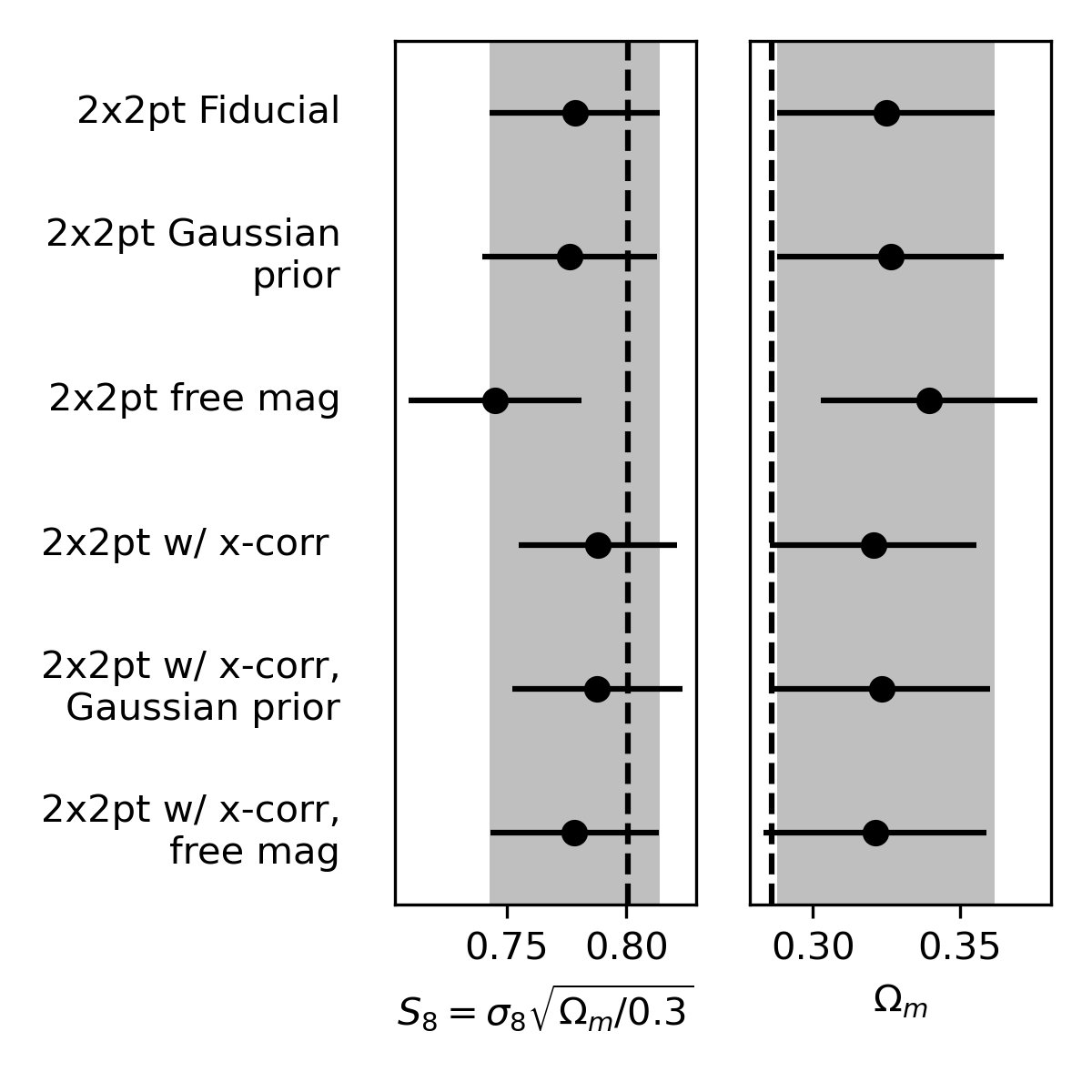}
\includegraphics[width=0.59\linewidth]{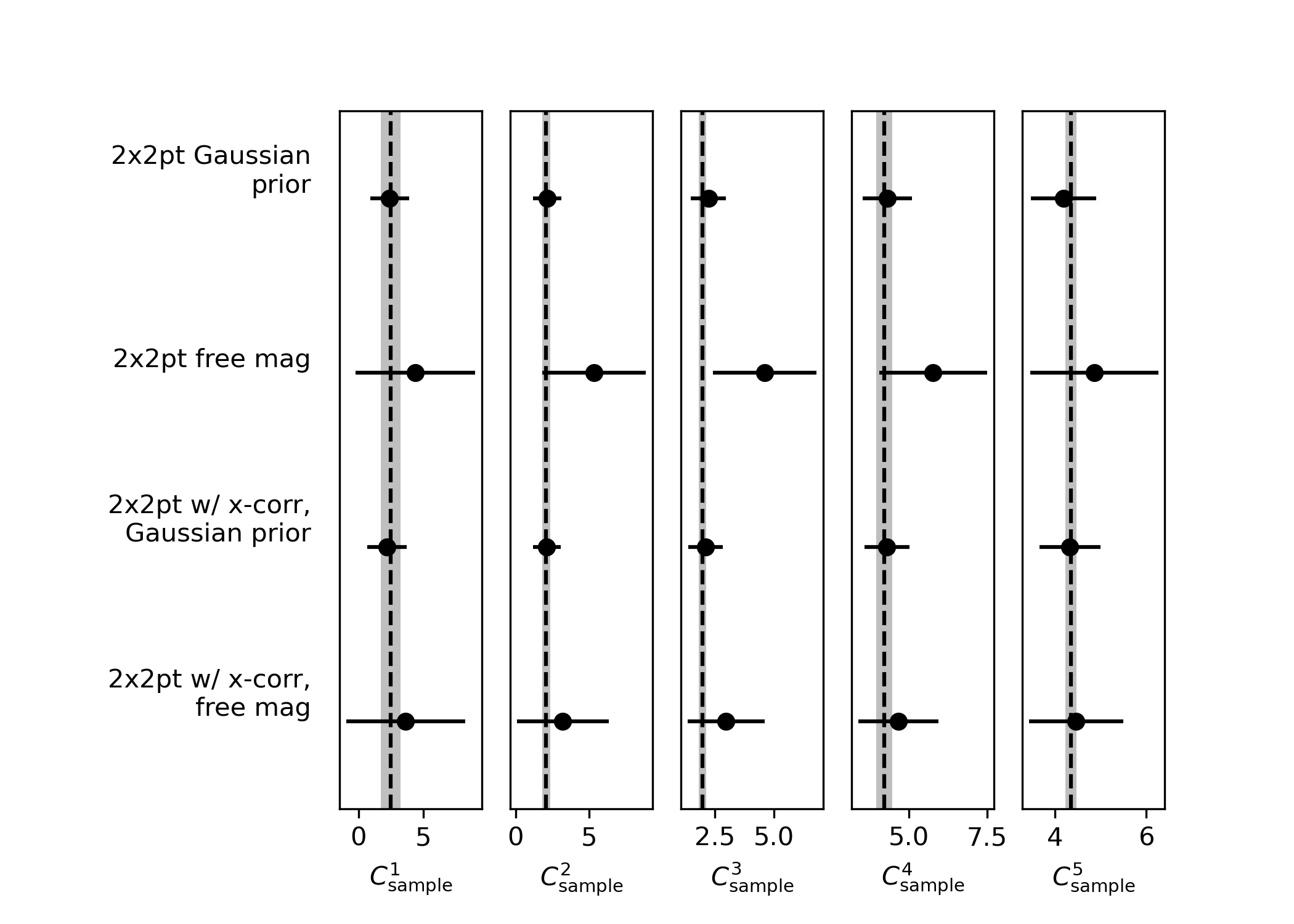}
\caption{Constraints from running our parameter inference pipeline on the Buzzard simulations (using the \redmagic\ sample). Left: Posterior mean and $1\sigma$ uncertainties on the recovered $S_8$ and \Om\, with the true (i.e.\ input to the simulations) values indicated by the black dashed line). Right: Constraints on the magnification coefficients \csample, for each lens redshift bin. In both, we show constraints from various different analysis variations w.r.t.\ our fiducial setup, which did not vary the magnification coefficients (see \sect{sec:buzzard_validation} for further discussion).}
\label{fig:buzzard_s8}
\end{figure*}

The right-hand panel of \fig{fig:buzzard_s8} compares the recovered magnification coefficients from the Buzzard analysis at fixed cosmology, with and without including \wtheta\ cross-correlations in the datavector. These constraints are compared to the values directly measured in the simulations as described in \sect{sec:magsims}. We see that the coefficients are well recovered (for the uniform prior case as well as the more trivial Gaussian prior case), with better constraining power when \wtheta\ cross-correlations are included. We are confident therefore that our modeling of the lens magnification signal in the two point functions matches our estimates in \sect{sec:magsims} for the Buzzard simulations.

We further test recovery of the true cosmology for the same 3 choices of prior on the magnification coefficients used in Sec.~\ref{sec:simresults}: a wide flat prior, a Gaussian prior from the statistical \balrog uncertainties on the data, and fixed to their true values. We use the \balrog uncertainties for the Gaussian prior to mimic the prior size in the real data analysis. We find consistent recovery of cosmological constraints as for the theory model datavector case in \sect{sec:simresults} i.e. offsets with the true cosmology appear consistent with being due to prior volume/projection effects. When a wide, uniform prior is used on the magnification coefficients \csample, these volume effects appear to increase (see third line of \fig{fig:buzzard_s8} left panel). 

The inclusion  of cross-correlations in \wtheta\ appears to reduce these volume effects, producing with $S_8$ and $\Om$ constraints almost identical to those from the fiducial analysis (see sixth line of \fig{fig:buzzard_s8} left panel). 

\section{DES Year 3 Results}
\label{sec:results}

In this section we present cosmology constraints from the DES Y3 2$\times$2pt data, investigating the impact of magnification-related analysis choices on cosmological constraints, and test the internal consistency of the magnification modeling. 

\subsection{Impact of  Magnification Priors}

\begin{figure*}
    \centering
    \begin{tabular}{ccc}
    \subfigure{\includegraphics[width=0.3\linewidth]{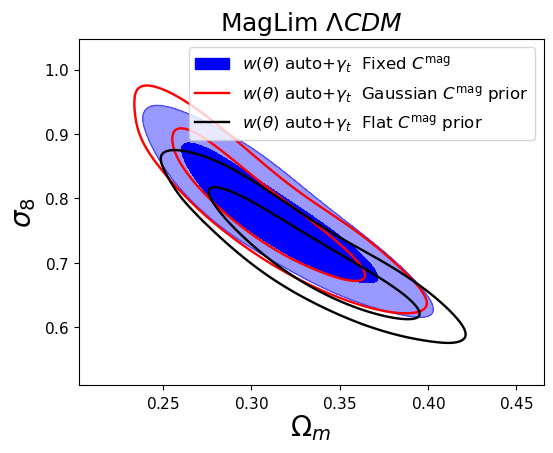}}
    &
    \subfigure{\includegraphics[width=0.3\linewidth]{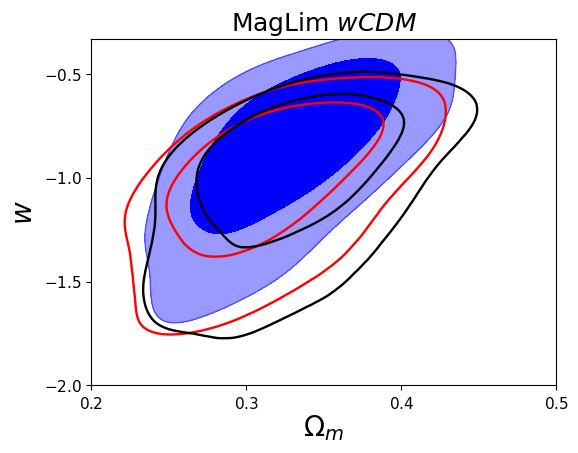}}
    &
    \subfigure{\includegraphics[width=0.3\linewidth]{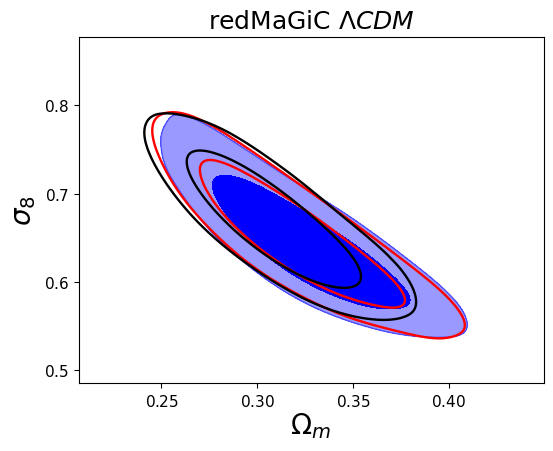}}
    \end{tabular}
    
    \caption{ Un-blind cosmology constraints on data with different magnification priors. Each panel shows cosmology constraints with the magnification coefficients $C_{total}$, fixed at the best fit value from \balrog, with a Gaussian prior from \balrog, and with a wide flat prior. left: \lcdm \maglim 2$\times$2pt constraints, centre: \wcdm \maglim 2$\times$2pt constraints, right: \lcdm \redmagic\ 2$\times$2pt constraints. }
    \label{fig:result_cosmo}
\end{figure*}

In \fig{fig:result_cosmo} we show the impact of allowing $\csample$ to vary with the same three sets of priors used in Sections ~\ref{sec:simresults} and ~\ref{sec:buzzard_validation}: fixed to their fiducial values, Gaussian priors with widths based on the \balrog\ uncertainties (including the systematic uncertainties we assigned as described in \sect{sec:cflux_discussion}), and uniform priors. We show results for \lcdm\ and \wcdm\ models for the \maglim\ sample, and for \lcdm\ only in the \redmagic\ sample (as \maglim is the default sample in the 3$\times$2pt analysis \citep{y3-3x2ptkp}, who suggested the potential presence of observational systematics in the \redmagic\ measurements). The cosmological constraints are broadly robust to changing the priors on the magnification coefficients. 

We find that allowing the magnification parameters to vary within the Gaussian prior set by the \balrog simulations does not notably change the cosmological constraints. However, when allowing \csample to vary with a wide uninformative prior, the \maglim\ \lcdm\ $S_8$ constraint is lower than the fixed case by ${\sim}1.2\sigma$. This behaviour is consistent with that seen when analysing the simulated datavectors in Secs.~\ref{sec:simresults} and \ref{sec:buzzard_validation}; where projection/volume effects resulted in a low $S_8$ constraint ($-1.4\sigma$ for a noiseless \maglim data vector) when the magnification parameters were varied freely. The improvement in goodness-of-fit indicated by the PPD is only modest (0.014 to 0.046), and therefore does not allow us to make definitive statements about whether free magnification coefficients are required by the data. While there may be other unmodelled systematic effects that complicate the picture in the case of the real data analysis (we discuss this further in \sect{sec:csampleresults}), we believe the similar size and direction of these shifts means it is reasonable to ascribe it to projection/volume effects. Thus we should be careful in interpreting these constraints; these projection/volume effects imply that when allowing the magnification coefficients to vary, without supplying any extra data, we enter a regime where we are in some sense trying to fit too many parameters to our data.

We discuss the degeneracy of \csample with other parameters, including intrinsic alignment amplitude, in Appendix \ref{sec:appendix_parameters_ia_gt}.

We discuss the \csample\ posteriors from these runs later in \sect{sec:csampleresults}.

\subsection{Clustering Cross-Correlations}

\begin{figure*}
    \subfigure{\includegraphics[width=0.75\linewidth]{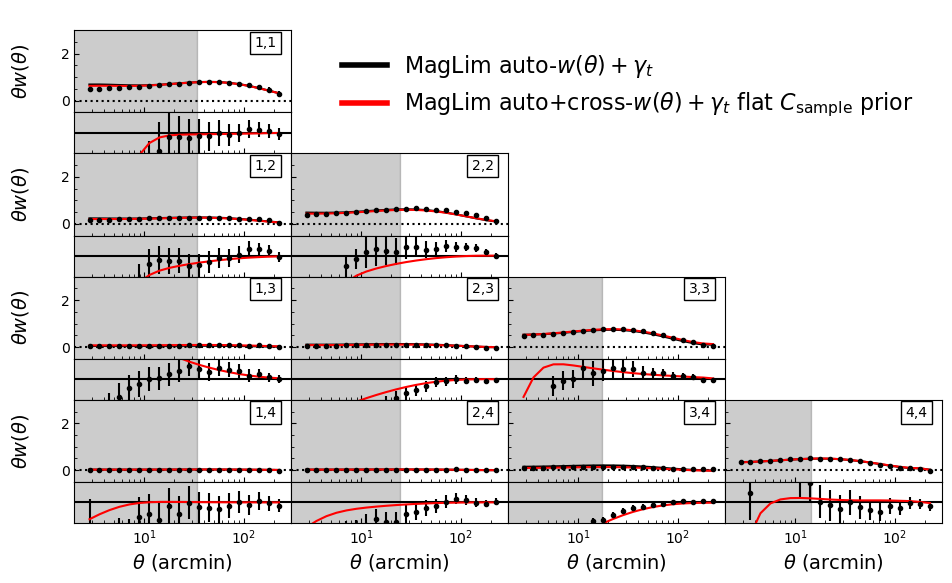}}\quad
    \subfigure{\includegraphics[width=0.9\linewidth]{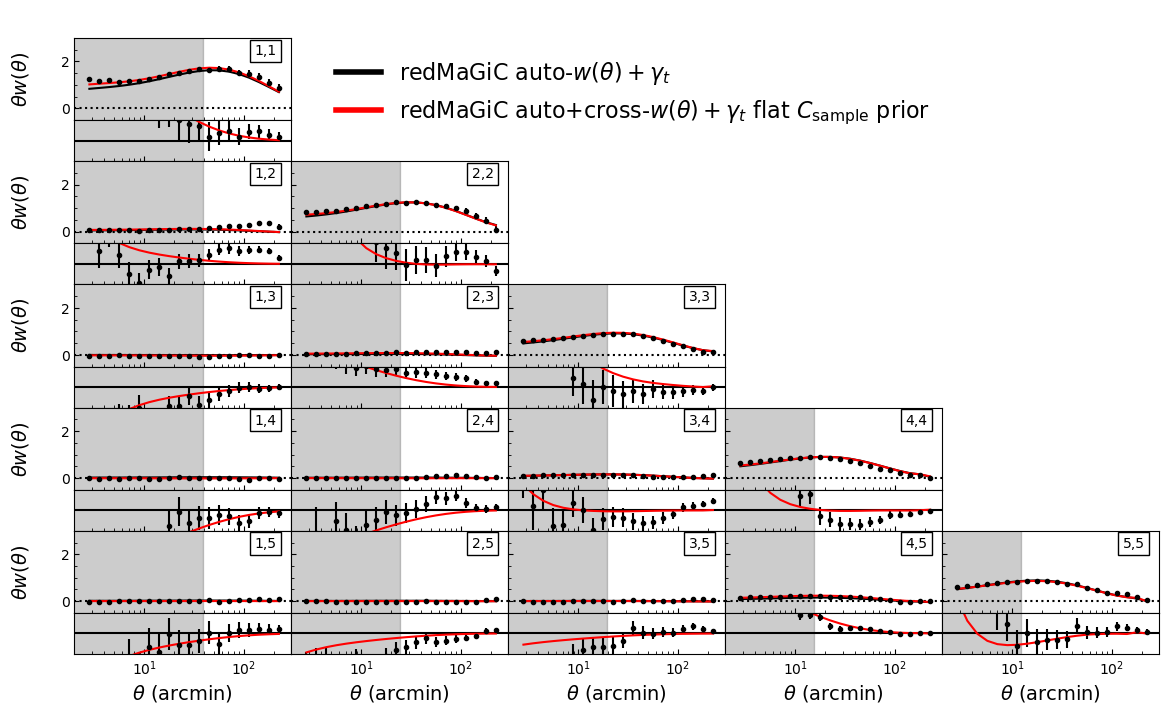}}
    \caption{The measured clustering signal, $\wtheta$ for the \maglim sample (top), and \redmagic sample (bottom), including cross-correlations between redshift bins. The lower panels in each block show the difference between the data and the best fit theory from the fiducial \lcdm analysis (which fixed magnification coefficients, and used only the auto-correlations of $w(\theta)$). Also shown in red is the prediction based on the best-fit parameters when including the cross-correlations between redshift bins in the fit, and allowing the magnification coefficients to vary. 
    For \maglim, the prediction based on the best-fit from the fiducial analysis is a poor fit to the cross-correlation measurements, with a $\chi^2$ of 130.7 for the 59 cross-correlation datapoints. The red line shows significant improvement, with a $\chi^2$ of 63.7. Note that we still exclude redshift bins 5 and 6 for \maglim, consistent with the fiducial 3$\times$2pt analysis. For \redmagic\, the $\chi^2$ of the $\wtheta$ cross-correlations improves from 167.6 to 144.7 (for 96 data points) when the cross-correlations are included in the fit and magnification coefficients allowed to vary. So while there is significant improvement in the fit, it remains quite poor, probably suggesting the presence of further unaccounted-for systematics.
    }
    \label{fig:cross_correlation}
\end{figure*}

\begin{figure*}
    \centering
    \begin{tabular}{ccc}
    \subfigure{\includegraphics[width=0.3\linewidth]{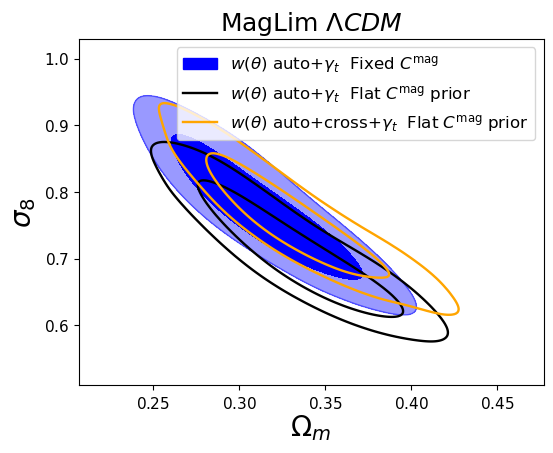}}
    &
    \subfigure{\includegraphics[width=0.3\linewidth]{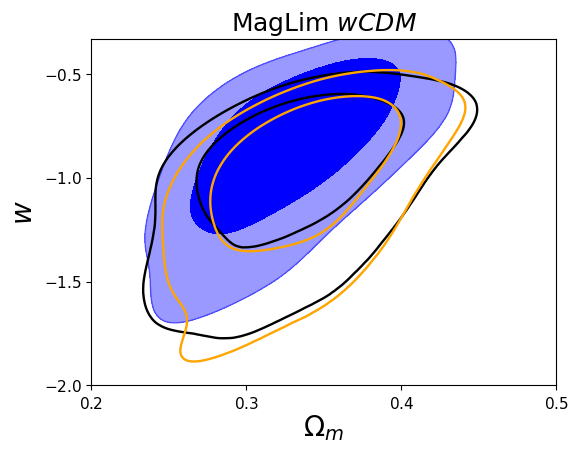}}
    &
    \subfigure{\includegraphics[width=0.3\linewidth]{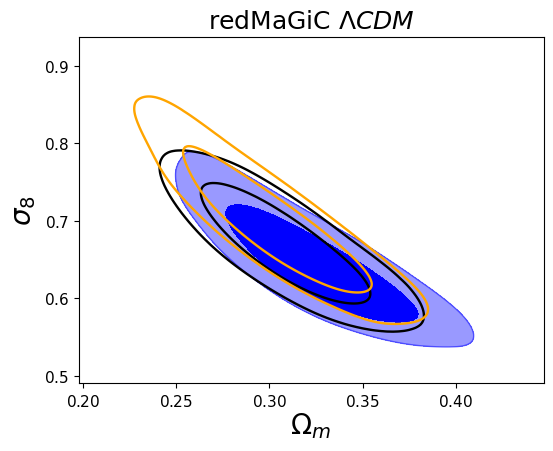}}
    \end{tabular}
    
    \caption{Cosmology constraints on data with and without including $w(\theta)$ cross-correlations. Each panel shows cosmology constraints with auto-only fixed magnification coefficients, auto-only free magnification coefficients, and auto+cross free magnification coefficients. left: \lcdm \maglim 2$\times$2pt constraints, centre: \wcdm \maglim 2$\times$2pt constraints, right: \lcdm \redmagic\ 2$\times$2pt constraints. }
    \label{fig:result_cosmo_xcorr}
\end{figure*}

We now explore the addition of the clustering  cross-correlation measurements between lens redshift bins, with the intention of better constraining the magnification signal. 

In \fig{fig:cross_correlation} we show the $\wtheta$ cross-correlations (between redshift bins) measured on the real data compared to the best-fit 2$\times$2pt from the fiducial analysis that only uses auto-correlations in the fit. The covariance matrix was computed using the Cosmolike package \citep{krause17,fang20} following the procedure in \cite{y3-generalmethods}. 
The inclusion of the cross-correlations is sensitive to both the magnification bias and the accuracy of the tails of the lens sample redshift distributions. Note that unless stated otherwise, we follow the main \threebytwo\ analysis and exclude the two highest redshift \maglim\ bins.

The measured \maglim cross correlations are systematically lower than the expectation from the fiducial analysis, which has fixed magnification coefficients; the 59 cross-correlation datapoints have a $\chi^2$ of 130.7. This indicates that either the magnification coefficients estimated using \balrog\ are inaccurate, or some other systematic is present, for example, these cross-correlations are likely to be more sensitive to the tails of the lens redshift distributions than when using only auto-correlations. When the cross-correlations are included in the fit and the magnification coefficients are allowed to vary this $\chi^2$ reduces to 63.7. 

For \redmagic\ we also see significant deviations of the measured cross-correlations from the prediction based on the fiducial analysis, with a $\chi^2$ of the cross-correlation data of 167.6 for 96 data-points, reducing to 144.7 when cross-correlations are included in the fit and magnification is allowed to vary. So while there is significant improvement in the fit, it remains quite poor, probably suggesting the presence of further unaccounted-for systematics.



We then investigate whether including $\wtheta$ cross-correlations and allowing the magnification coefficients to vary i) affects the cosmological parameter inference ii) improves the model fit to the $\wtheta$ cross-correlations.
\fig{fig:result_cosmo_xcorr} demonstrates that the cosmological constraints are robust to the inclusion of $\wtheta$. Allowing the $\csample$ values to vary in the analysis naturally loses some constraining power (although perhaps due to the presence of projection/volume effects, this is not apparent in the width of the resulting $S_8$ constraint, and may manifest instead as a shift in the $S_8$ constraint). Adding the clustering between bins allows us to regain some of the lost information. Including the clustering between bins gives cosmology constrains quite consistent with the fiducial analysis, with the mean of the $S_8$ posterior increasing by 0.2$\sigma$ (0.9$\sigma$) for \maglim\ (\redmagic). 

In \tab{tab:chi2_table}, we show the the p-value for the posterior predictive distribution (PPD) for each cosmology MCMC chain (see \citealt{doux21} for details). These values can be interpreted as a goodness-of-fit metric, indicating how likely the particular realization of the Y3 data (for the full \twobytwo\ data vector as well as $\gammat$ and $w(\theta)$ separately) is, given the assumed cosmological model. When allowing magnification coefficients to vary, we see a moderate improvement in the PPD values, for both lens galaxy samples. When additionally including $\wtheta$ cross-correlations, we see no significant change in the p-value, suggesting at least that our modelling framework can be applied to the $\wtheta$ cross-correlation at reasonably successfully. 

\subsection{Magnification coefficient (\csample) constraints}\label{sec:csampleresults}

\begin{figure*}
    \centering
    \subfigure{\includegraphics[width=0.9\linewidth]{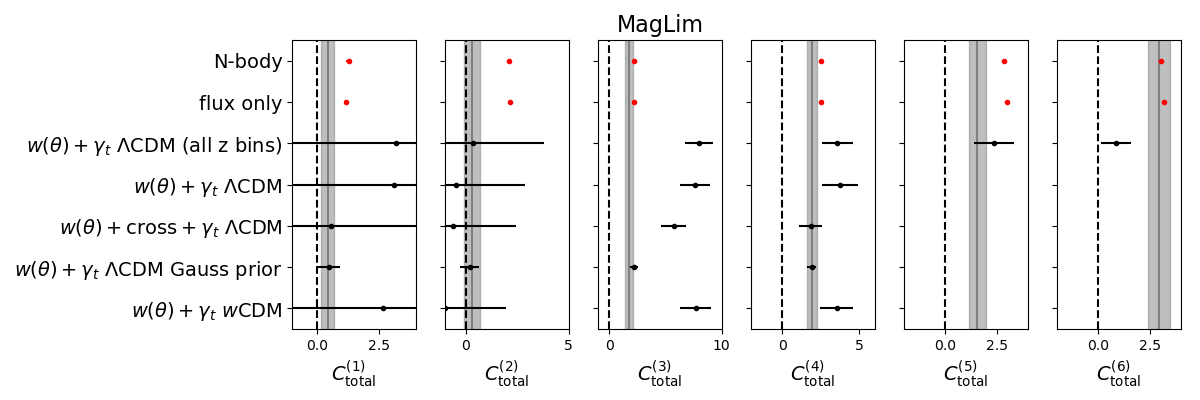}}\quad
    \subfigure{\includegraphics[width=0.8\linewidth]{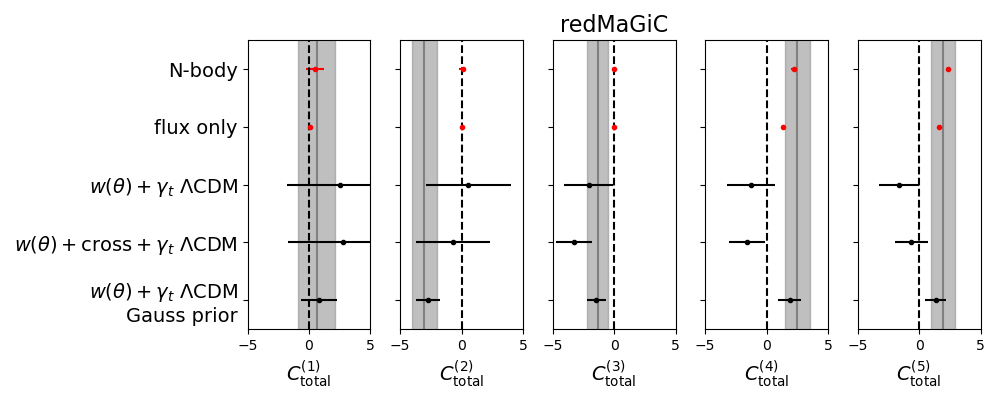}}
    
    \caption{ Unblinded posterior constraints on magnification coefficients $C_{\rm total}$ in the \maglim sample (top), and \redmagic sample (bottom). The vertical grey band is the estimate from the \balrog image simulations. The red points were estimated on the data prior to the cosmology analysis. The black points are $68\%$ posterior bounds from the analysis of galaxy clustering and galaxy-galaxy lensing. }
    \label{fig:result_mag}
\end{figure*}

In \fig{fig:result_mag} we show the posterior distribution of the $\csample$ in each redshift bin compared to the estimates from \balrog i.e. those that we fixed $\csample$ to in the fiducial analysis. The $\csample$ posterior in the third redshift bin of the \maglim sample is significantly larger than any of the estimates from \balrog, N-body simulations or the flux perturbations. When adding the cross-correlations between lens bins, this constraint moves closer to the estimated values, though remains significantly higher. Given the good agreement between the prior estimates in this redshift bin, it is likely the magnification parameters are capturing some other systematic in the DES data that biases the clustering measurements. 

The fourth \maglim redshift bin posteriors are in good agreement with the prior estimates, especially when $\wtheta$ cross-correlations are included. Given their low redshift and therefore weak magnification signal, the constraining power on the magnification coefficients for the lowest two redshift bins is too weak to make a useful comparison.

We also analyse a 2$\times$2pt data vector using all six \maglim redshift bins, keeping all other analysis choices the same as in the fiducial analysis. We find the posteriors on the first four \csample parameters are consistent with the four bin run. The fifth bin is consistent with expectation from \balrog. However, the sixth bin is around 3$\sigma$ lower than expectation. Given that these redshift bins were excluded from the fiducial analysis due to the likely presence of (not-well-understood) measurement systematics, we cannot ascribe a specific cause to this discrepancy.

For the \redmagic sample, we find good agreement between the posteriors and prior estimates in the first three redshift bins (although again the constraining power for the first two bins is weak). For the two highest redshift bins, the 2$\times$2pt posteriors favor lower $\csample$ values at 2-3$\sigma$. Again, this may be an indication of the magnification coefficients' sensitivity to other observational systematics (see discussion in \citealt{y3-3x2ptkp, y3-galaxyclustering, y3-2x2ptbiasmodelling}). 

\begingroup
\setlength{\tabcolsep}{12pt} 
\begin{table*}
    \centering
    \begin{tabular}{llllll}
        \hline 
        \multicolumn{6}{c}{\textbf{\maglim}}   \\ \hline 
        Data        & Model& Mag prior  & PPD 2$\times$2pt & PPD $\gamma_{t}$ & PPD $w(\theta)$\\ \hline
        2$\times$2pt         & \lcdm & fixed      & 0.014   &  0.015   &  0.182 \\
        2$\times$2pt         & \lcdm & Gaussian   & 0.038   &  0.047   &  0.340 \\
        2$\times$2pt         & \lcdm & flat       & 0.046   &  0.056   &  0.311 \\
        2$\times$2pt + cross & \lcdm & flat       & 0.052   &  0.063   &  0.345 \\
        2$\times$2pt         & \wcdm & fixed      & 0.024   &  0.029   &  0.208 \\
        2$\times$2pt         & \wcdm & Gaussian   & 0.054   &  0.048   &  0.366 \\
        2$\times$2pt         & \wcdm & flat       & 0.053   &  0.050   &  0.346 \\
        2$\times$2pt + cross & \wcdm & Flat       & 0.040   &  0.038   &  0.211 \\
        2$\times$2pt (6 bins)& \lcdm & Flat       & 0.046   &  0.042   &  0.231 \\
        \hline
        \multicolumn{6}{c}{\textbf{\redmagic}}   \\ \hline 
        Data        & Model& Mag prior  & PPD 2$\times$2pt & PPD $\gamma_{t}$ & PPD $w(\theta)$\\ \hline
        2$\times$2pt         & \lcdm & fixed      & 0.025   &  0.028   &  0.171 \\
        2$\times$2pt         & \lcdm & Gaussian   & 0.056   &  0.107   &  0.098 \\
        2$\times$2pt         & \lcdm & flat       & 0.095   &  0.150   &  0.127 \\
        2$\times$2pt + cross & \lcdm & flat       & 0.093   &  0.134   &  0.126 \\
        \hline
    \end{tabular}
    \caption{The p-value for the posterior predictive distribution (PPD) for each cosmology MCMC chain (see \citealt{doux21} for details). 
    These values indicate how likely the particular realization of the Y3 data (for the full \twobytwo\ data vector as well as $\gammat$ and $w(\theta)$ separately) is, given the assumed cosmological model. The p-value is therefore a measure of the goodness of fit.  
    For both lens samples, there is a moderate improvement in p-value when allowing magnification parameters to vary. There is no significant change to the p-value when cross-correlations are added.
    }
    \label{tab:chi2_table}
\end{table*}
\endgroup

In general, when allowing the magnification coefficients to vary freely in the analysis we do not always recover values expected from the prior estimates described in \sect{sec:method}. This could be due to missing effects in the image simulations used to estimate the magnification coefficients that were not captured by any of the methods in \sect{sec:method}. It could also be driven by observational systematics in the data vector that are degenerate with the magnification signal. These potential systematics are further explored in \citet{y3-3x2ptkp,y3-2x2ptaltlensresults, y3-2x2ptbiasmodelling, y3-galaxyclustering}. 
\section{Conclusions}
\label{sec:conclusions}

This analysis has studied the impact of lensing magnification of the lens sample in the Dark Energy Survey Year 3 cosmological analysis,  which combines galaxy clustering with galaxy weak lensing \citep{y3-3x2ptkp}. We estimate the amplitude of the magnification coefficients via several methods prior to analysis of the measured datavector, test the model assumptions of the fiducial analysis, and infer cosmological parameter constraints with different modelling choices related to lens magnification. 

We estimate the amplitude of the magnification coefficients from the  realistic \balrog\ image simulations \citep{y3-balrog}, by injecting fake objects into real DES images and testing the response of the number density of selected objects to magnifying the input objects. We compare these estimates with simplified estimates from directly perturbing fluxes in the real data, and from mock catalog simulations,  and find some deviations, especially at  at intermediate redshifts. These differences may be driven by the additional  effects included in the \balrog\ image simulations, such as the impact on galaxy size and photometric redshift estimates.


By running variants on the fiducial DES Y3 analysis, on both simulated and real data, we demonstrate that the fiducial analysis choice, where magnification coefficients were fixed at the best-fit estimates from the image simulations, does not bias the cosmological inference relative to imposing an informative prior that accounts for uncertainty in those  estimates. We constrain cosmological parameters in \lcdm\ and wCDM with different sets of priors on magnification coefficients and find the cosmology constraints on the data to behave as expected from the simulation, where changes in the $S_8$ constraint less than $1\sigma$ can be induced by incorrect magnification modelling or projection effects from a model that is too flexible. 

We demonstrate the usefulness of including the clustering cross-correlations between lens redshift bins to better constrain the magnification coefficients, and therefore also the cosmological parameters. In our simulated analyses in \sect{sec:simresults}, we show that on the Buzzard simulation measurement, including the $\wtheta$ cross-correlations allows one to vary freely the magnification coefficients, without incurring a cost in constraining power on $S_8$ with respect to the fiducial analysis (which fixed the magnification coefficients and did not include   $\wtheta$ cross-correlations). This opens up the possibility of a more conservative analysis with minimal assumptions on the  magnification coefficient values. It also points to the potential for extracting significant cosmological information from the $\wtheta$ cross-correlations if the magnification coefficient values can be calibrated from simulations, as we've attempted here.

We constrain the magnification coefficients $\csample$ from the two-point functions themselves, and find some of the coefficients to be inconsistent with their prior estimates from both the \balrog image simulations, and the alternative methods. While this could be caused by incorrect input coefficients, we believe the extreme values are more likely to be induced by other unmodelled systematics in the DES \twobytwo \ data. 
Despite this preference for unexpected values of the magnification coefficients, when freeing the magnification parameters we observe shifts in cosmological parameters that are very similar to those biases expected from projection/prior volume effects, as observed on simulated where the modelling is perfect and no systematics are present. It is therefore unlikely that the fiducial DES Year 3 approach is incurring large ( $\sim>1\sigma$) biases in cosmological parameters due to degeneracies with magnification parameters that have been fixed to incorrect values, since in this case we would see cosmological parameter shift when freeing magnification coefficients that are inconsistent across data and simulations. Rather, it is likely that i) there are projection/prior volume effects that are introduced by freeing the magnification coefficients ii) the extreme and unexpected inferred values of magnfication parameters are due to unmodelled systematics in the data. In addition, the same systematic effects could also be biasing the cosmological parameters in both the fixed and free magnification coefficients cases, and requires further  investigation before we can claim the DES Year 3 constraints are fully robust.
One unmodelled systematic that could be degenerate with magnification is the impact of dust extinction from galaxies that trace structures between us and the lens galaxies. This extinction could also produce changes in the observed number density of galaxies that is correlated with foreground large-scale structure, although unlike lensing, its impact on galaxy fluxes is chromatic \citep{Menard2010}.
The \csample inconsistency is interesting in the context of the much studied high-redshift inconsistency between clustering and galaxy-galaxy lensing in DES Year 3 \citep{y3-2x2ptbiasmodelling, y3-2x2ptaltlensresults, y3-3x2ptkp}. However, because some of the \csample posteriors are significantly discrepant with all of the methods explored in this paper, we also consider the possibility that the magnification signal is partially degenerate with this inconsistency, rather than the cause. 

The results in this paper demonstrate the DES 2$\times$2pt cosmology results are broadly robust to the pre-unblinding choices of magnification prior, and changes in cosmological parameters follow the expected small shifts seen on simulations. The magnification signal itself is detectable in the 2$\times$2pt data vector, but can be sensitive to systematics.

\section*{Acknowledgments}

Funding for the DES Projects has been provided by the DOE and NSF(USA), MEC/MICINN/MINECO(Spain), STFC(UK), HEFCE(UK). NCSA(UIUC), KICP(U. Chicago), CCAPP(Ohio State), 
MIFPA(Texas A\&M), CNPQ, FAPERJ, FINEP (Brazil), DFG(Germany) and the Collaborating Institutions in the Dark Energy Survey.

The Collaborating Institutions are Argonne Lab, UC Santa Cruz, University of Cambridge, CIEMAT-Madrid, University of Chicago, University College London, 
DES-Brazil Consortium, University of Edinburgh, ETH Z{\"u}rich, Fermilab, University of Illinois, ICE (IEEC-CSIC), IFAE Barcelona, Lawrence Berkeley Lab, 
LMU M{\"u}nchen and the associated Excellence Cluster Universe, University of Michigan, NFS's NOIRLab, University of Nottingham, Ohio State University, University of 
Pennsylvania, University of Portsmouth, SLAC National Lab, Stanford University, University of Sussex, Texas A\&M University, and the OzDES Membership Consortium.

Based in part on observations at Cerro Tololo Inter-American Observatory at NSF's NOIRLab (NOIRLab Prop. ID 2012B-0001; PI: J. Frieman), which is managed by the Association of Universities for Research in Astronomy (AURA) under a cooperative agreement with the National Science Foundation.

The DES Data Management System is supported by the NSF under Grant Numbers AST-1138766 and AST-1536171. 
The DES participants from Spanish institutions are partially supported by MICINN under grants ESP2017-89838, PGC2018-094773, PGC2018-102021, SEV-2016-0588, SEV-2016-0597, and MDM-2015-0509, some of which include ERDF funds from the European Union. IFAE is partially funded by the CERCA program of the Generalitat de Catalunya.
Research leading to these results has received funding from the European Research
Council under the European Union's Seventh Framework Program (FP7/2007-2013) including ERC grant agreements 240672, 291329, and 306478.
We  acknowledge support from the Brazilian Instituto Nacional de Ci\^encia
e Tecnologia (INCT) do e-Universo (CNPq grant 465376/2014-2).

This manuscript has been authored by Fermi Research Alliance, LLC under Contract No. DE-AC02-07CH11359 with the U.S. Department of Energy, Office of Science, Office of High Energy Physics. 

\section*{Affiliations}
\noindent

\noindent
{\it
$^{1}$ Center for Cosmology and Astro-Particle Physics, The Ohio State University, Columbus, OH 43210, USA\\
$^{2}$ Department of Physics, The Ohio State University, Columbus, OH 43210, USA\\
$^{3}$ Department of Physics and Astronomy, University of Waterloo, 200 University Ave W, Waterloo, ON N2L 3G1, Canada\\
$^{4}$ Department of Applied Mathematics and Theoretical Physics, University of Cambridge, Cambridge CB3 0WA, UK\\
$^{5}$ Jet Propulsion Laboratory, California Institute of Technology, 4800 Oak Grove Dr., Pasadena, CA 91109, USA\\
$^{6}$ Department of Astronomy and Astrophysics, University of Chicago, Chicago, IL 60637, USA\\
$^{7}$ Kavli Institute for Cosmological Physics, University of Chicago, Chicago, IL 60637, USA\\
$^{8}$ Kavli Institute for Particle Astrophysics \& Cosmology, P. O. Box 2450, Stanford University, Stanford, CA 94305, USA\\
$^{9}$ SLAC National Accelerator Laboratory, Menlo Park, CA 94025, USA\\
$^{10}$ Centro de Investigaciones Energ\'eticas, Medioambientales y Tecnol\'ogicas (CIEMAT), Madrid, Spain\\
$^{11}$ Fermi National Accelerator Laboratory, P. O. Box 500, Batavia, IL 60510, USA\\
$^{12}$ Jodrell Bank Center for Astrophysics, School of Physics and Astronomy, University of Manchester, Oxford Road, Manchester, M13 9PL, UK\\
$^{13}$ California Institute of Technology, 1200 East California Blvd, MC 249-17, Pasadena, CA 91125, USA\\
$^{14}$ Argonne National Laboratory, 9700 South Cass Avenue, Lemont, IL 60439, USA\\
$^{15}$ Department of Physics, University of Michigan, Ann Arbor, MI 48109, USA\\
$^{16}$ Institute of Astronomy, University of Cambridge, Madingley Road, Cambridge CB3 0HA, UK\\
$^{17}$ Kavli Institute for Cosmology, University of Cambridge, Madingley Road, Cambridge CB3 0HA, UK\\
$^{18}$ Institute for Astronomy, University of Hawai'i, 2680 Woodlawn Drive, Honolulu, HI 96822, USA\\
$^{19}$ Physics Department, 2320 Chamberlin Hall, University of Wisconsin-Madison, 1150 University Avenue Madison, WI  53706-1390\\
$^{20}$ Department of Physics and Astronomy, University of Pennsylvania, Philadelphia, PA 19104, USA\\
$^{21}$ Department of Physics, Northeastern University, Boston, MA 02115, USA\\
$^{22}$ Instituto de F\'{i}sica Te\'orica, Universidade Estadual Paulista, S\~ao Paulo, Brazil\\
$^{23}$ Laborat\'orio Interinstitucional de e-Astronomia - LIneA, Rua Gal. Jos\'e Cristino 77, Rio de Janeiro, RJ - 20921-400, Brazil\\
$^{24}$ Department of Physics, Carnegie Mellon University, Pittsburgh, Pennsylvania 15312, USA\\
$^{25}$ Instituto de Astrofisica de Canarias, E-38205 La Laguna, Tenerife, Spain\\
$^{26}$ Universidad de La Laguna, Dpto. Astrofísica, E-38206 La Laguna, Tenerife, Spain\\
$^{27}$ Center for Astrophysical Surveys, National Center for Supercomputing Applications, 1205 West Clark St., Urbana, IL 61801, USA\\
$^{28}$ Department of Astronomy, University of Illinois at Urbana-Champaign, 1002 W. Green Street, Urbana, IL 61801, USA\\
$^{29}$ Physics Department, William Jewell College, Liberty, MO, 64068\\
$^{30}$ Department of Physics, Duke University Durham, NC 27708, USA\\
$^{31}$ Institut d'Estudis Espacials de Catalunya (IEEC), 08034 Barcelona, Spain\\
$^{32}$ Institute of Space Sciences (ICE, CSIC),  Campus UAB, Carrer de Can Magrans, s/n,  08193 Barcelona, Spain\\
$^{33}$ Lawrence Berkeley National Laboratory, 1 Cyclotron Road, Berkeley, CA 94720, USA\\
$^{34}$ NSF AI Planning Institute for Physics of the Future, Carnegie Mellon University, Pittsburgh, PA 15213, USA\\
$^{35}$ \\
$^{36}$ Department of Astronomy/Steward Observatory, University of Arizona, 933 North Cherry Avenue, Tucson, AZ 85721-0065, USA\\
$^{37}$ Department of Physics \& Astronomy, University College London, Gower Street, London, WC1E 6BT, UK\\
$^{38}$ Department of Astronomy, University of California, Berkeley,  501 Campbell Hall, Berkeley, CA 94720, USA\\
$^{39}$ Institut de F\'{\i}sica d'Altes Energies (IFAE), The Barcelona Institute of Science and Technology, Campus UAB, 08193 Bellaterra (Barcelona) Spain\\
$^{40}$ University Observatory, Faculty of Physics, Ludwig-Maximilians-Universit\"at, Scheinerstr. 1, 81679 Munich, Germany\\
$^{41}$ School of Physics and Astronomy, Cardiff University, CF24 3AA, UK\\
$^{42}$ Department of Astronomy, University of Geneva, ch. d'\'Ecogia 16, CH-1290 Versoix, Switzerland\\
$^{43}$ Department of Physics, University of Arizona, Tucson, AZ 85721, USA\\
$^{44}$ Department of Physics and Astronomy, Pevensey Building, University of Sussex, Brighton, BN1 9QH, UK\\
$^{45}$ Instituto de Astrof\'{\i}sica e Ci\^{e}ncias do Espa\c{c}o, Faculdade de Ci\^{e}ncias, Universidade de Lisboa, 1769-016 Lisboa, Portugal\\
$^{46}$ Perimeter Institute for Theoretical Physics, 31 Caroline St. North, Waterloo, ON N2L 2Y5, Canada\\
$^{47}$ Department of Physics, Stanford University, 382 Via Pueblo Mall, Stanford, CA 94305, USA\\
$^{48}$ Instituto de F\'isica Gleb Wataghin, Universidade Estadual de Campinas, 13083-859, Campinas, SP, Brazil\\
$^{49}$ Kavli Institute for the Physics and Mathematics of the Universe (WPI), UTIAS, The University of Tokyo, Kashiwa, Chiba 277-8583, Japan\\
$^{50}$ Institute for Astronomy, University of Edinburgh, Edinburgh EH9 3HJ, UK\\
$^{51}$ ICTP South American Institute for Fundamental Research\\ Instituto de F\'{\i}sica Te\'orica, Universidade Estadual Paulista, S\~ao Paulo, Brazil\\
$^{52}$ Brookhaven National Laboratory, Bldg 510, Upton, NY 11973, USA\\
$^{53}$ Department of Physics and Astronomy, Stony Brook University, Stony Brook, NY 11794, USA\\
$^{54}$ D\'{e}partement de Physique Th\'{e}orique and Center for Astroparticle Physics, Universit\'{e} de Gen\`{e}ve, 24 quai Ernest Ansermet, CH-1211 Geneva, Switzerland\\
$^{55}$ Excellence Cluster Origins, Boltzmannstr.\ 2, 85748 Garching, Germany\\
$^{56}$ Max Planck Institute for Extraterrestrial Physics, Giessenbachstrasse, 85748 Garching, Germany\\
$^{57}$ Universit\"ats-Sternwarte, Fakult\"at f\"ur Physik, Ludwig-Maximilians Universit\"at M\"unchen, Scheinerstr. 1, 81679 M\"unchen, Germany\\
$^{58}$ George P. and Cynthia Woods Mitchell Institute for Fundamental Physics and Astronomy, and Department of Physics and Astronomy, Texas A\&M University, College Station, TX 77843,  USA\\
$^{59}$ Instituto de Fisica Teorica UAM/CSIC, Universidad Autonoma de Madrid, 28049 Madrid, Spain\\
$^{60}$ Institute of Cosmology and Gravitation, University of Portsmouth, Portsmouth, PO1 3FX, UK\\
$^{61}$ CNRS, UMR 7095, Institut d'Astrophysique de Paris, F-75014, Paris, France\\
$^{62}$ Sorbonne Universit\'es, UPMC Univ Paris 06, UMR 7095, Institut d'Astrophysique de Paris, F-75014, Paris, France\\
$^{63}$ Department of Astronomy and Astrophysics, University of Toronto, 50 St. George Street, Toronto ON, M5S 3H4, Canada\\
}

\section*{Data Availability}

All Dark Energy Survey data used in this work is available through the cosmological data release on the Dark Energy Survey Data Management (DESDM) system \url{https://des.ncsa.illinois.edu/releases/y3a2}.




\bibliographystyle{mnras_2author}
\bibliography{main,y3kp} 

\begin{thebibliography}{}
\makeatletter
\relax
\def\mn@urlcharsother{\let\do\@makeother \do\$\do\&\do\#\do\^\do\_\do\%\do\~}
\def\mn@doi{\begingroup\mn@urlcharsother \@ifnextchar [ {\mn@doi@}
  {\mn@doi@[]}}
\def\mn@doi@[#1]#2{\def\@tempa{#1}\ifx\@tempa\@empty \href
  {http://dx.doi.org/#2} {doi:#2}\else \href {http://dx.doi.org/#2} {#1}\fi
  \endgroup}
\def\mn@eprint#1#2{\mn@eprint@#1:#2::\@nil}
\def\mn@eprint@arXiv#1{\href {http://arxiv.org/abs/#1} {{\tt arXiv:#1}}}
\def\mn@eprint@dblp#1{\href {http://dblp.uni-trier.de/rec/bibtex/#1.xml}
  {dblp:#1}}
\def\mn@eprint@#1:#2:#3:#4\@nil{\def\@tempa {#1}\def\@tempb {#2}\def\@tempc
  {#3}\ifx \@tempc \@empty \let \@tempc \@tempb \let \@tempb \@tempa \fi \ifx
  \@tempb \@empty \def\@tempb {arXiv}\fi \@ifundefined
  {mn@eprint@\@tempb}{\@tempb:\@tempc}{\expandafter \expandafter \csname
  mn@eprint@\@tempb\endcsname \expandafter{\@tempc}}}

\bibitem[\protect\citeauthoryear{Abbott \& Aguena et~al.,}{Abbott
  et~al.}{2022}]{y3-3x2ptkp}
Abbott T. M.~C.,  et~al. 2022, \mn@doi [Phys. Rev. D]
  {10.1103/PhysRevD.105.023520}, 105, 023520

\bibitem[\protect\citeauthoryear{{Amon} \& {Gruen} et~al.,}{{Amon}
  et~al.}{2022}]{y3-cosmicshear1}
{Amon} A.,  et~al. 2022, \mn@doi [\prd] {10.1103/PhysRevD.105.023514}, \href
  {https://ui.adsabs.harvard.edu/abs/2022PhRvD.105b3514A} {105, 023514}

\bibitem[\protect\citeauthoryear{{Bartelmann} \& {Schneider}}{{Bartelmann} \&
  {Schneider}}{2001}]{bartelmann01}
{Bartelmann} M.,  {Schneider} P.,  2001, \mn@doi [\physrep]
  {10.1016/S0370-1573(00)00082-X}, \href
  {https://ui.adsabs.harvard.edu/abs/2001PhR...340..291B} {340, 291}

\bibitem[\protect\citeauthoryear{{Becker}}{{Becker}}{2013}]{Becker2013}
{Becker} M.~R.,  2013, \mn@doi [\mnras] {10.1093/mnras/stt1352}, \href
  {https://ui.adsabs.harvard.edu/abs/2013MNRAS.435..115B} {435, 115}

\bibitem[\protect\citeauthoryear{{Bernstein}}{{Bernstein}}{2009}]{bernstein09}
{Bernstein} G.~M.,  2009, \mn@doi [\apj] {10.1088/0004-637X/695/1/652}, \href
  {https://ui.adsabs.harvard.edu/abs/2009ApJ...695..652B} {695, 652}

\bibitem[\protect\citeauthoryear{{Bertin} \& {Arnouts}}{{Bertin} \&
  {Arnouts}}{1996}]{bertin96}
{Bertin} E.,  {Arnouts} S.,  1996, \mn@doi [\aaps] {10.1051/aas:1996164}, \href
  {https://ui.adsabs.harvard.edu/abs/1996A&AS..117..393B} {117, 393}

\bibitem[\protect\citeauthoryear{{Blanton} et~al.}{{Blanton}
  et~al.}{2003}]{2003ApJ...592..819B}
{Blanton} M.~R.,  et~al., 2003, \mn@doi [\apj] {10.1086/375776}, \href
  {https://ui.adsabs.harvard.edu/abs/2003ApJ...592..819B} {592, 819}

\bibitem[\protect\citeauthoryear{{Blazek} \& {MacCrann} et~al.,}{{Blazek}
  et~al.}{2019}]{tatt}
{Blazek} J.~A.,  et~al. 2019, \mn@doi [\prd] {10.1103/PhysRevD.100.103506},
  \href {https://ui.adsabs.harvard.edu/abs/2019PhRvD.100j3506B} {100, 103506}

\bibitem[\protect\citeauthoryear{{Brammer}, {van Dokkum}  \& {Coppi}}{{Brammer}
  et~al.}{2008}]{easyphotoz_brammer}
{Brammer} G.~B.,  {van Dokkum} P.~G.,   {Coppi} P.,  2008, \mn@doi [\apj]
  {10.1086/591786}, \href
  {https://ui.adsabs.harvard.edu/abs/2008ApJ...686.1503B} {686, 1503}

\bibitem[\protect\citeauthoryear{{Carretero} \& {Castander}
  et~al.,}{{Carretero} et~al.}{2015}]{2015MNRAS.447..646C}
{Carretero} J.,  et~al. 2015, \mn@doi [\mnras] {10.1093/mnras/stu2402}, \href
  {https://ui.adsabs.harvard.edu/abs/2015MNRAS.447..646C} {447, 646}

\bibitem[\protect\citeauthoryear{{Crocce}, {Pueblas}  \&
  {Scoccimarro}}{{Crocce} et~al.}{2006}]{Crocce2005}
{Crocce} M.,  {Pueblas} S.,   {Scoccimarro} R.,  2006, \mn@doi [\mnras]
  {10.1111/j.1365-2966.2006.11040.x}, \href
  {https://ui.adsabs.harvard.edu/abs/2006MNRAS.373..369C} {373, 369}

\bibitem[\protect\citeauthoryear{{Crocce} \& {Castander} et~al.,}{{Crocce}
  et~al.}{2015}]{2015MNRAS.453.1513C}
{Crocce} M.,  et~al. 2015, \mn@doi [\mnras] {10.1093/mnras/stv1708}, \href
  {https://ui.adsabs.harvard.edu/abs/2015MNRAS.453.1513C} {453, 1513}

\bibitem[\protect\citeauthoryear{{De Vicente}, {S{\'a}nchez}  \&
  {Sevilla-Noarbe}}{{De Vicente} et~al.}{2016}]{DNF}
{De Vicente} J.,  {S{\'a}nchez} E.,   {Sevilla-Noarbe} I.,  2016, \mn@doi
  [\mnras] {10.1093/mnras/stw857}, \href
  {https://ui.adsabs.harvard.edu/abs/2016MNRAS.459.3078D} {459, 3078}

\bibitem[\protect\citeauthoryear{{DeRose}, {Becker}  \& {Wechsler}}{{DeRose}
  et~al.}{2021}]{DeRose2021a}
{DeRose} J.,  {Becker} M.,   {Wechsler} R.,  2021, in prep.

\bibitem[\protect\citeauthoryear{DeRose \& Wechsler et~al.,}{DeRose
  et~al.}{2022}]{y3-simvalidation}
DeRose J.,  et~al. 2022, \mn@doi [Phys. Rev. D] {10.1103/PhysRevD.105.123520},
  105, 123520

\bibitem[\protect\citeauthoryear{{Deshpande} \& {Kitching}}{{Deshpande} \&
  {Kitching}}{2020}]{deshpande_2020}
{Deshpande} A.~C.,  {Kitching} T.~D.,  2020, \mn@doi [\prd]
  {10.1103/PhysRevD.101.103531}, \href
  {https://ui.adsabs.harvard.edu/abs/2020PhRvD.101j3531D} {101, 103531}

\bibitem[\protect\citeauthoryear{{Doux} \& {Baxter} et~al.,}{{Doux}
  et~al.}{2021}]{doux21}
{Doux} C.,  et~al. 2021, \mn@doi [\mnras] {10.1093/mnras/stab526}, \href
  {https://ui.adsabs.harvard.edu/abs/2021MNRAS.503.2688D} {503, 2688}

\bibitem[\protect\citeauthoryear{Duncan \& Harnois-D{\'{e}}raps et~al.,}{Duncan
  et~al.}{2022}]{Duncan_2022}
Duncan C. A.~J.,  et~al. 2022, \mn@doi [\mnras] {10.1093/mnras/stac1809}, 515,
  1130

\bibitem[\protect\citeauthoryear{{Everett} \& {Yanny} et~al.,}{{Everett}
  et~al.}{2022}]{y3-balrog}
{Everett} S.,  et~al. 2022, \mn@doi [\apjs] {10.3847/1538-4365/ac26c1}, \href
  {https://ui.adsabs.harvard.edu/abs/2022ApJS..258...15E} {258, 15}

\bibitem[\protect\citeauthoryear{{Fang}, {Eifler}  \& {Krause}}{{Fang}
  et~al.}{2020a}]{fang20}
{Fang} X.,  {Eifler} T.,   {Krause} E.,  2020a, \mn@doi [\mnras]
  {10.1093/mnras/staa1726}, \href
  {https://ui.adsabs.harvard.edu/abs/2020MNRAS.497.2699F} {497, 2699}

\bibitem[\protect\citeauthoryear{{Fang} \& {Krause} et~al.,}{{Fang}
  et~al.}{2020b}]{Fang_nonlimber}
{Fang} X.,  et~al. 2020b, \mn@doi [\jcap] {10.1088/1475-7516/2020/05/010},
  \href {https://ui.adsabs.harvard.edu/abs/2020JCAP...05..010F} {2020, 010}

\bibitem[\protect\citeauthoryear{{Flaugher} et~al.}{{Flaugher}
  et~al.}{2015}]{Flaugher:2015}
{Flaugher} B.,  et~al., 2015, \mn@doi [\aj] {10.1088/0004-6256/150/5/150},
  \href {https://ui.adsabs.harvard.edu/abs/2015AJ....150..150F} {150, 150}

\bibitem[\protect\citeauthoryear{{Fosalba} \& {Gazta{\~n}aga}
  et~al.,}{{Fosalba} et~al.}{2015a}]{2015MNRAS.447.1319F}
{Fosalba} P.,  et~al. 2015a, \mn@doi [\mnras] {10.1093/mnras/stu2464}, \href
  {https://ui.adsabs.harvard.edu/abs/2015MNRAS.447.1319F} {447, 1319}

\bibitem[\protect\citeauthoryear{{Fosalba} \& {Crocce} et~al.,}{{Fosalba}
  et~al.}{2015b}]{2015MNRAS.448.2987F}
{Fosalba} P.,  et~al. 2015b, \mn@doi [\mnras] {10.1093/mnras/stv138}, \href
  {https://ui.adsabs.harvard.edu/abs/2015MNRAS.448.2987F} {448, 2987}

\bibitem[\protect\citeauthoryear{{Garcia-Fernandez} et~al.}{{Garcia-Fernandez}
  et~al.}{2016}]{sv-magnification}
{Garcia-Fernandez} M.,  et~al., 2016, arXiv e-prints, \href
  {https://ui.adsabs.harvard.edu/abs/2016arXiv161110326G} {p. arXiv:1611.10326}

\bibitem[\protect\citeauthoryear{Gatti, Giannini  et~al.}{Gatti
  et~al.}{2020}]{y3-sourcewz}
Gatti M.,  Giannini G.,   et~al., 2020, Submitted to MNRAS

\bibitem[\protect\citeauthoryear{Gatti, Sheldon  et~al.}{Gatti
  et~al.}{2021}]{y3-shapecatalog}
Gatti M.,  Sheldon E.,   et~al., 2021, \mn@doi [\mnras]
  {10.1093/mnras/stab918}, \href
  {https://ui.adsabs.harvard.edu/abs/2021MNRAS.504.4312G} {504, 4312}

\bibitem[\protect\citeauthoryear{{G{\'o}rski} et~al.}{{G{\'o}rski}
  et~al.}{2005}]{healpix}
{G{\'o}rski} K.~M.,  et~al., 2005, \mn@doi [\apj] {10.1086/427976}, \href
  {http://adsabs.harvard.edu/abs/2005ApJ...622..759G} {622, 759}

\bibitem[\protect\citeauthoryear{Hartley, Choi  et~al.}{Hartley
  et~al.}{2022}]{y3-deepfields}
Hartley W.~G.,  Choi A.,   et~al., 2022, \mn@doi [\mnras]
  {10.1093/mnras/stab3055}, \href
  {https://ui.adsabs.harvard.edu/abs/2022MNRAS.509.3547H} {509, 3547}

\bibitem[\protect\citeauthoryear{{Hu} \& {Jain}}{{Hu} \&
  {Jain}}{2004}]{hujain2004}
{Hu} W.,  {Jain} B.,  2004, \mn@doi [\prd] {10.1103/PhysRevD.70.043009}, \href
  {https://ui.adsabs.harvard.edu/abs/2004PhRvD..70d3009H} {70, 043009}

\bibitem[\protect\citeauthoryear{{Huff} \& {Mandelbaum}}{{Huff} \&
  {Mandelbaum}}{2017}]{metacalibration}
{Huff} E.,  {Mandelbaum} R.,  2017, arXiv e-prints, \href
  {https://ui.adsabs.harvard.edu/abs/2017arXiv170202600H} {p. arXiv:1702.02600}

\bibitem[\protect\citeauthoryear{{Ilbert} et~al.}{{Ilbert}
  et~al.}{2009}]{2009ApJ...690.1236I}
{Ilbert} O.,  et~al., 2009, \mn@doi [\apj] {10.1088/0004-637X/690/2/1236},
  \href {https://ui.adsabs.harvard.edu/abs/2009ApJ...690.1236I} {690, 1236}

\bibitem[\protect\citeauthoryear{{Joachimi} \& {Bridle}}{{Joachimi} \&
  {Bridle}}{2010}]{joachimi2010}
{Joachimi} B.,  {Bridle} S.~L.,  2010, \mn@doi [\aap]
  {10.1051/0004-6361/200913657}, \href
  {https://ui.adsabs.harvard.edu/abs/2010A&A...523A...1J} {523, A1}

\bibitem[\protect\citeauthoryear{Kaiser}{Kaiser}{1987}]{10.1093/mnras/227.1.1}
Kaiser N.,  1987, \mn@doi [\mnras] {10.1093/mnras/227.1.1}, 227, 1

\bibitem[\protect\citeauthoryear{{Kamionkowski}, {Kosowsky}  \&
  {Stebbins}}{{Kamionkowski} et~al.}{1997}]{kamionkowski97}
{Kamionkowski} M.,  {Kosowsky} A.,   {Stebbins} A.,  1997, \mn@doi [\prd]
  {10.1103/PhysRevD.55.7368}, \href
  {https://ui.adsabs.harvard.edu/abs/1997PhRvD..55.7368K} {55, 7368}

\bibitem[\protect\citeauthoryear{{Krause} \& {Eifler}}{{Krause} \&
  {Eifler}}{2017}]{krause17}
{Krause} E.,  {Eifler} T.,  2017, \mn@doi [\mnras] {10.1093/mnras/stx1261},
  \href {https://ui.adsabs.harvard.edu/abs/2017MNRAS.470.2100K} {470, 2100}

\bibitem[\protect\citeauthoryear{{Krause} \& {Fang} et~al.,}{{Krause}
  et~al.}{2021}]{y3-generalmethods}
{Krause} E.,  et~al. 2021, arXiv e-prints, \href
  {https://ui.adsabs.harvard.edu/abs/2021arXiv210513548K} {p. arXiv:2105.13548}

\bibitem[\protect\citeauthoryear{Lewis, Challinor  \& Lasenby}{Lewis
  et~al.}{2000}]{Lewis2000}
Lewis A.,  Challinor A.,   Lasenby A.,  2000, \mn@doi [\apj] {10.1086/309179},
  538, 473

\bibitem[\protect\citeauthoryear{{Lorenz}, {Alonso}  \& {Ferreira}}{{Lorenz}
  et~al.}{2018}]{lorenz_2018}
{Lorenz} C.~S.,  {Alonso} D.,   {Ferreira} P.~G.,  2018, \mn@doi [\prd]
  {10.1103/PhysRevD.97.023537}, \href
  {https://ui.adsabs.harvard.edu/abs/2018PhRvD..97b3537L} {97, 023537}

\bibitem[\protect\citeauthoryear{{MacCrann} \& {Becker} et~al.,}{{MacCrann}
  et~al.}{2022}]{y3-imagesims}
{MacCrann} N.,  et~al. 2022, \mn@doi [\mnras] {10.1093/mnras/stab2870}, \href
  {https://ui.adsabs.harvard.edu/abs/2022MNRAS.509.3371M} {509, 3371}

\bibitem[\protect\citeauthoryear{{Mahony} \& {Fortuna} et~al.,}{{Mahony}
  et~al.}{2022}]{mahoney22}
{Mahony} C.,  et~al. 2022, \mn@doi [\mnras] {10.1093/mnras/stac872}, \href
  {https://ui.adsabs.harvard.edu/abs/2022MNRAS.513.1210M} {513, 1210}

\bibitem[\protect\citeauthoryear{{Mandelbaum} \& {Slosar} et~al.,}{{Mandelbaum}
  et~al.}{2013}]{mandelbaum2013}
{Mandelbaum} R.,  et~al. 2013, \mn@doi [\mnras] {10.1093/mnras/stt572}, \href
  {https://ui.adsabs.harvard.edu/abs/2013MNRAS.432.1544M} {432, 1544}

\bibitem[\protect\citeauthoryear{{M{\'e}nard} \& {Scranton}
  et~al.,}{{M{\'e}nard} et~al.}{2010}]{Menard2010}
{M{\'e}nard} B.,  et~al. 2010, \mn@doi [\mnras]
  {10.1111/j.1365-2966.2010.16486.x}, \href
  {https://ui.adsabs.harvard.edu/abs/2010MNRAS.405.1025M} {405, 1025}

\bibitem[\protect\citeauthoryear{{Morganson} et~al.}{{Morganson}
  et~al.}{2018}]{Morganson:2018}
{Morganson} E.,  et~al., 2018, \mn@doi [\pasp] {10.1088/1538-3873/aab4ef},
  \href {https://ui.adsabs.harvard.edu/abs/2018PASP..130g4501M} {130, 074501}

\bibitem[\protect\citeauthoryear{{Myles}, {Alarcon}  et~al.}{{Myles}
  et~al.}{2021}]{y3-sompz}
{Myles} J.,  {Alarcon} A.,   et~al., 2021, \mn@doi [\mnras]
  {10.1093/mnras/stab1515}, \href
  {https://ui.adsabs.harvard.edu/abs/2021MNRAS.505.4249M} {505, 4249}

\bibitem[\protect\citeauthoryear{Pandey \& Krause et~al.,}{Pandey
  et~al.}{2022}]{y3-2x2ptbiasmodelling}
Pandey S.,  et~al. 2022, \mn@doi [Phys. Rev. D] {10.1103/PhysRevD.106.043520},
  106, 043520

\bibitem[\protect\citeauthoryear{{Porredon} \& {Crocce} et~al.,}{{Porredon}
  et~al.}{2021a}]{y3-2x2ptaltlensresults}
{Porredon} A.,  et~al. 2021a, arXiv e-prints, \href
  {https://ui.adsabs.harvard.edu/abs/2021arXiv210513546P} {p. arXiv:2105.13546}

\bibitem[\protect\citeauthoryear{Porredon et~al.}{Porredon
  et~al.}{2021b}]{y3-2x2maglimforecast}
Porredon A.,  et~al., 2021b, \mn@doi [Phys. Rev. D]
  {10.1103/PhysRevD.103.043503}, 103, 043503

\bibitem[\protect\citeauthoryear{{Prat} \& {Blazek} et~al.,}{{Prat}
  et~al.}{2022}]{y3-gglensing}
{Prat} J.,  et~al. 2022, \mn@doi [\prd] {10.1103/PhysRevD.105.083528}, \href
  {https://ui.adsabs.harvard.edu/abs/2022PhRvD.105h3528P} {105, 083528}

\bibitem[\protect\citeauthoryear{{Rodr{\'\i}guez-Monroy} \& {Weaverdyck}
  et~al.,}{{Rodr{\'\i}guez-Monroy} et~al.}{2022}]{y3-galaxyclustering}
{Rodr{\'\i}guez-Monroy} M.,  et~al. 2022, \mn@doi [\mnras]
  {10.1093/mnras/stac104}, \href
  {https://ui.adsabs.harvard.edu/abs/2022MNRAS.511.2665R} {511, 2665}

\bibitem[\protect\citeauthoryear{{Rowe} et~al.}{{Rowe} et~al.}{2015}]{galsim}
{Rowe} B.~T.~P.,  et~al., 2015, \mn@doi [Astronomy and Computing]
  {10.1016/j.ascom.2015.02.002}, \href
  {https://ui.adsabs.harvard.edu/abs/2015A&C....10..121R} {10, 121}

\bibitem[\protect\citeauthoryear{{Rozo} et~al.}{{Rozo}
  et~al.}{2016}]{sv-redmagic}
{Rozo} E.,  et~al., 2016, \mn@doi [\mnras] {10.1093/mnras/stw1281}, \href
  {http://adsabs.harvard.edu/abs/2016MNRAS.461.1431R} {461, 1431}

\bibitem[\protect\citeauthoryear{{Rykoff} et~al.}{{Rykoff}
  et~al.}{2014}]{rykoff14}
{Rykoff} E.~S.,  et~al., 2014, \mn@doi [\apj] {10.1088/0004-637X/785/2/104},
  \href {https://ui.adsabs.harvard.edu/abs/2014ApJ...785..104R} {785, 104}

\bibitem[\protect\citeauthoryear{{Rykoff} et~al.}{{Rykoff}
  et~al.}{2016}]{rykoff16}
{Rykoff} E.~S.,  et~al., 2016, \mn@doi [\apjs] {10.3847/0067-0049/224/1/1},
  \href {https://ui.adsabs.harvard.edu/abs/2016ApJS..224....1R} {224, 1}

\bibitem[\protect\citeauthoryear{Scranton et~al.}{Scranton
  et~al.}{2005}]{Scranton:2005ci}
Scranton R.,  et~al., 2005, \mn@doi [Astrophys. J.] {10.1086/431358}, 633, 589

\bibitem[\protect\citeauthoryear{{Secco}, {Samuroff}  et~al.}{{Secco}
  et~al.}{2022}]{y3-cosmicshear2}
{Secco} L.~F.,  {Samuroff} S.,   et~al., 2022, \mn@doi [\prd]
  {10.1103/PhysRevD.105.023515}, \href
  {https://ui.adsabs.harvard.edu/abs/2022PhRvD.105b3515S} {105, 023515}

\bibitem[\protect\citeauthoryear{{Sevilla-Noarbe} \& {Bechtol}
  et~al.,}{{Sevilla-Noarbe} et~al.}{2021}]{y3-gold}
{Sevilla-Noarbe} I.,  et~al. 2021, \mn@doi [\apjs] {10.3847/1538-4365/abeb66},
  \href {https://ui.adsabs.harvard.edu/abs/2021ApJS..254...24S} {254, 24}

\bibitem[\protect\citeauthoryear{Springel}{Springel}{2005}]{Springel2005}
Springel V.,  2005, \mn@doi [\mnras] {10.1111/j.1365-2966.2005.09655.x}, 364,
  1105

\bibitem[\protect\citeauthoryear{{Stebbins}}{{Stebbins}}{1996}]{stebbins96}
{Stebbins} A.,  1996, arXiv e-prints, \href
  {https://ui.adsabs.harvard.edu/abs/1996astro.ph..9149S} {pp
  astro--ph/9609149}

\bibitem[\protect\citeauthoryear{{Thiele}, {Duncan}  \& {Alonso}}{{Thiele}
  et~al.}{2020}]{thiele_2020}
{Thiele} L.,  {Duncan} C. A.~J.,   {Alonso} D.,  2020, \mn@doi [\mnras]
  {10.1093/mnras/stz3103}, \href
  {https://ui.adsabs.harvard.edu/abs/2020MNRAS.491.1746T} {491, 1746}

\bibitem[\protect\citeauthoryear{{Wechsler} \& {DeRose} et~al.,}{{Wechsler}
  et~al.}{2022}]{Wechsler2021}
{Wechsler} R.~H.,  et~al. 2022, \mn@doi [\apj] {10.3847/1538-4357/ac5b0a},
  \href {https://ui.adsabs.harvard.edu/abs/2022ApJ...931..145W} {931, 145}

\bibitem[\protect\citeauthoryear{Yoo \& Seljak}{Yoo \&
  Seljak}{2012}]{yooseljak12}
Yoo J.,  Seljak U.,  2012, \mn@doi [Phys. Rev. D] {10.1103/PhysRevD.86.083504},
  86, 083504

\bibitem[\protect\citeauthoryear{{Zehavi} et~al.}{{Zehavi}
  et~al.}{2011}]{2011ApJ...736...59Z}
{Zehavi} I.,  et~al., 2011, \mn@doi [\apj] {10.1088/0004-637X/736/1/59}, \href
  {https://ui.adsabs.harvard.edu/abs/2011ApJ...736...59Z} {736, 59}

\bibitem[\protect\citeauthoryear{{von Wietersheim-Kramsta} et~al.}{{von
  Wietersheim-Kramsta} et~al.}{2021}]{vonWietersheim-Kramsta:2021lid}
{von Wietersheim-Kramsta} M.,  et~al., 2021, \mn@doi [\mnras]
  {10.1093/mnras/stab1000}, \href
  {https://ui.adsabs.harvard.edu/abs/2021MNRAS.504.1452V} {504, 1452}

\makeatother
\end{thebibliography}

\appendix
\section{\Balrog Samples}
\label{app:balrog_plots}
\begin{figure*}
    \centering
    \subfigure{\includegraphics[width=\linewidth]{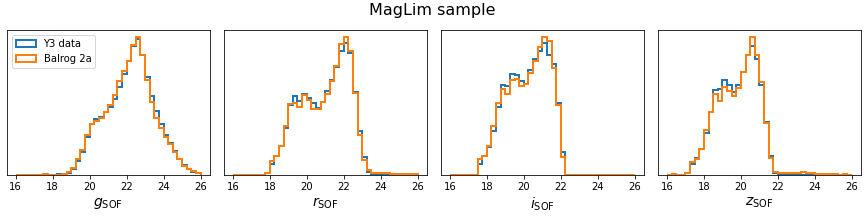}}
    \subfigure{\includegraphics[width=\linewidth]{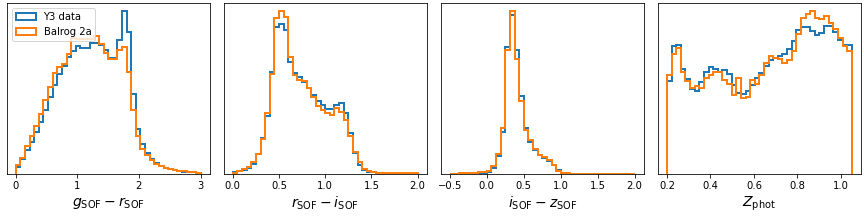}}
    \subfigure{\includegraphics[width=0.5\linewidth]{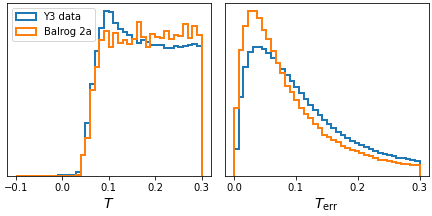}}
    
    \caption{ Histograms comparing quantities of the \maglim lens sample between the real data and \balrog. The quantities shown are SOF (single-object fitting) magnitudes and colors using bands $g$, $r$, $i$, and $z$; photometric redshift point estimate from the DNF algorithm used for selecting and binning the \maglim galaxies; and size estimates $T$ and $T_{\rm err}$ that were used for star-galaxy separation. The distribution of magnitude and color agree well between the samples, but the \balrog pipeline appears to underestimate the uncertainty on galaxy sizes $T_{\rm err}$. This difference is accounted for in the systematic uncertainty applied to \csample.     }
    \label{fig:data_vs_balrog_maglim}
\end{figure*}

\begin{figure*}
    \centering
    \subfigure{\includegraphics[width=\linewidth]{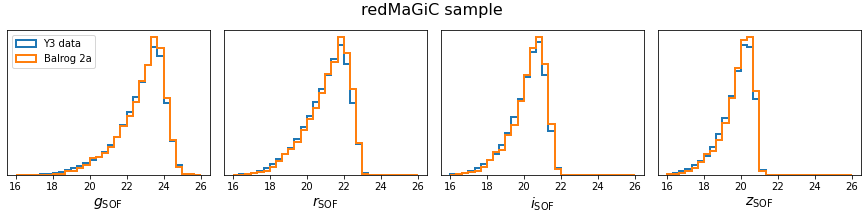}}
    \subfigure{\includegraphics[width=\linewidth]{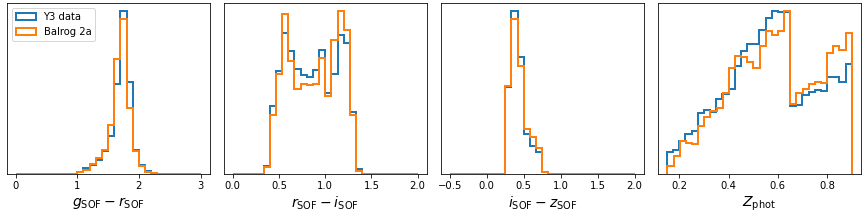}}
    \subfigure{\includegraphics[width=0.5\linewidth]{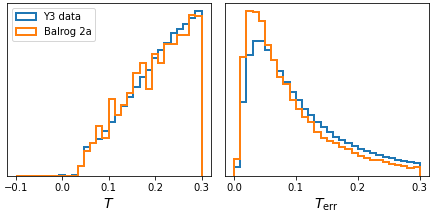}}
    
    \caption{ Same as \fig{fig:data_vs_balrog_maglim} but for the \redmagic sample. The photometric redshift shown is the $Z_{\redmagic}$ estimate calibrated from the same red sequence training as the data.    }
    \label{fig:data_vs_balrog_rm}
\end{figure*}

In this appendix we present a comparison between the real data and the \balrog lens samples used in this analysis. Histograms of the measured magnitude, color, photo-z and galaxy size are shown in \fig{fig:data_vs_balrog_maglim} for the \maglim sample and in \fig{fig:data_vs_balrog_rm} for \redmagic. We find good agreement between the color, magnitude and photometric redshift distributions between the two samples. There are some small differences in the distributions of galaxy size $T$ and size uncertainty $T_{\rm err}$. These quantities are used in the star galaxy separation cuts. In the \balrog sample, the photometric pipeline appears to generally find a slightly smaller size uncertainty than on the real data. Any differences between these two samples are accounted for by including a systematic error on \csample  equal to the difference between the flux only measurements in \balrog and the real data.

\section{Source Magnification}\label{app:source_mag}

As well as impacting the lens galaxies, magnification can also have an impact on the source sample by increasing the sampling of the shear field behind over-dense structures. This effect generally has a smaller impact on the 3$\times$2pt data vector than lens magnification as demonstrated for the DES year 3 analysis in \cite{y3-gglensing}. Source magnification also has a small impact estimations of source redshift uncertainties using the clustering redshift method as demonstrated for DES Year 3 in \cite*{y3-sourcewz}. 
In this appendix we show the results of applying the \balrog method to the source selection to obtain approximate \csample values for source galaxies. Since source magnification is not included in the baseline analysis modelling \cite{y3-generalmethods} for 3$\times$2pt, we present only the best fit \csample values output from the \balrog method, and do not perform the full error analysis. These values were used only to demonstrate insensitivity to source magnification.
The source sample was selected by running the metacalibration \cite{metacalibration} shape measurement pipeline on the balrog sample as described in \cite{y3-balrog} to produce a sample equivalent to the shape catalog described in \cite*{y3-shapecatalog}. 
In order to estimate the redshift dependence of the source magnification signal, we split the source sample into tomographic bins using the photometric redshift estimate of the \textit{deep field} object associated with each balrog injection. This photometric redshift point estimate came from the EAzY template fitting photo-z code \cite{easyphotoz_brammer}. The performance of this photo-z code on the DES deep fields is described in \cite*{y3-deepfields}. While these tomographic bins should not be considered the same as the photo-z binning of the wide field objects, it does demonstrate the redshift dependence of the \csample quantity in the redshift ranges relevant to the tomographic bins used in the main 3$\times$2pt analysis. The final SOMPZ redshift binning used in the final analysis \cite*{y3-sompz} was not available at the time these validation tests were performed for the Y3 analysis. 
The \csample value obtained for the non-tomographic source sample is $C_{\rm sample}^{\rm source \ notomo} = 1.90$. When split into tomographic bins using the deep field photo-z we obtain $\csample^{\rm source} = [0.67 , 1.37 , 1.986, 2.916]$. 

\section{Parameter Degeneracy in the free magnification analysis}
\label{sec:appendix_parameters_ia_gt}

In this appendix we further detail the parameter degeneracies in the 2$\times$2pt analysis when the magnification parameters \csample are allowed to be free. We aim to describe the features in the data that cause some \csample posteriors to disagree the the \balrog estimates in \fig{fig:result_mag}, and to describe the reasons for the small shifts in $\sigma_8$ in \fig{fig:result_cosmo}. 

\begin{figure*}
    \centering
    \subfigure{\includegraphics[width=\linewidth]{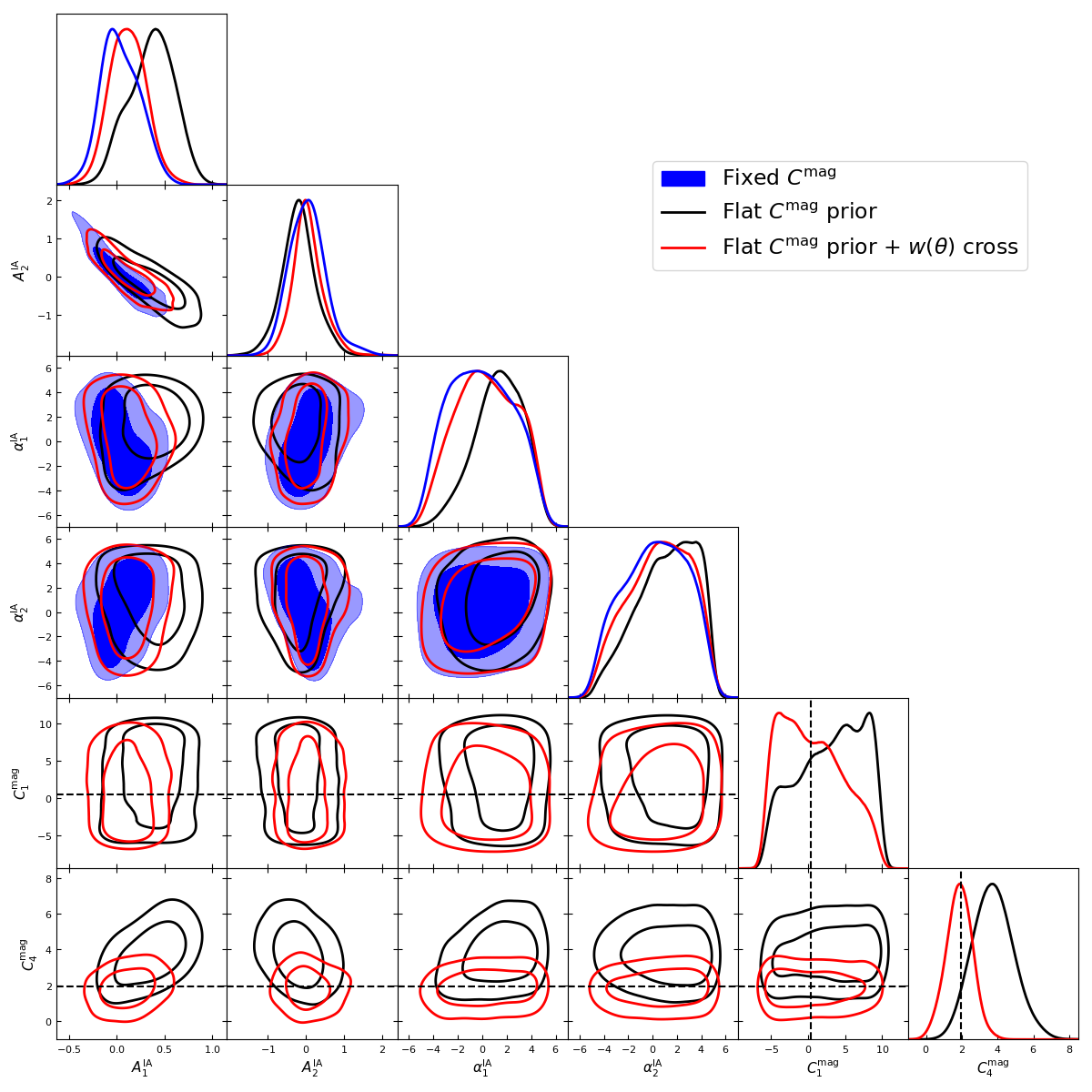}}
    
    \caption{ 2$\times$2pt \maglim constraints on  magnification coefficients and Intrinsic Alignment (IA) parameters in the TATT model. Results are shown for two magnification priors; a delta function at the \balrog best fits (dashed line) and the wide flat prior. Results are shown only for \csample in \maglim bins 1 and 4 for conciseness. We see a mild positive degeneracy between the IA amplitude and the constrained (higher-redshift) \csample. Freeing the magnification coefficients shifts the $A^{(IA)}_1$ posterior higher. We also show the posterior for the case with cross correlations between lens bins included in the fit, using a flat magnification prior (red). The cross-correlations constrain the high redshift \csample and shift the IA posteriors back towards the fixed case. Posteriors on \csample in the other tomographic bins were shown in Fig. \ref{fig:result_mag}}
    \label{fig:ia_params}
\end{figure*}

In \fig{fig:ia_params} we show the 2D parameter constraints of the Intrinsic Alignment parameters (using the TATT model \citep{tatt} which was the fiducial IA model in the DES year 3 analysis) and the \csample parameters from the lowest and highest \maglim redshift bins. We see some mild correlation between the IA amplitude parameter $A^{(IA)}_1$ and the constrained $\csample^4$. The other constrained \csample parameters show similar behaviour. This leads to a positive shift in the $A^{(IA)}_1$ posterior when magnification is freed. A higher $A^{(IA)}_1$ results in a lower $\sigma_8$, explaining the shift in \fig{fig:result_cosmo}.

Also in \fig{fig:ia_params}, we show the posteriors from a data vector including cross correlations between lens bins, with a flat prior on magnification parameters. One can see the cross-correlations move the magnification and IA posteriors towards the fixed magnification case, and break the degeneracy between magnification and IA. 

\begin{figure*}
    \centering
    \subfigure{\includegraphics[width=\linewidth]{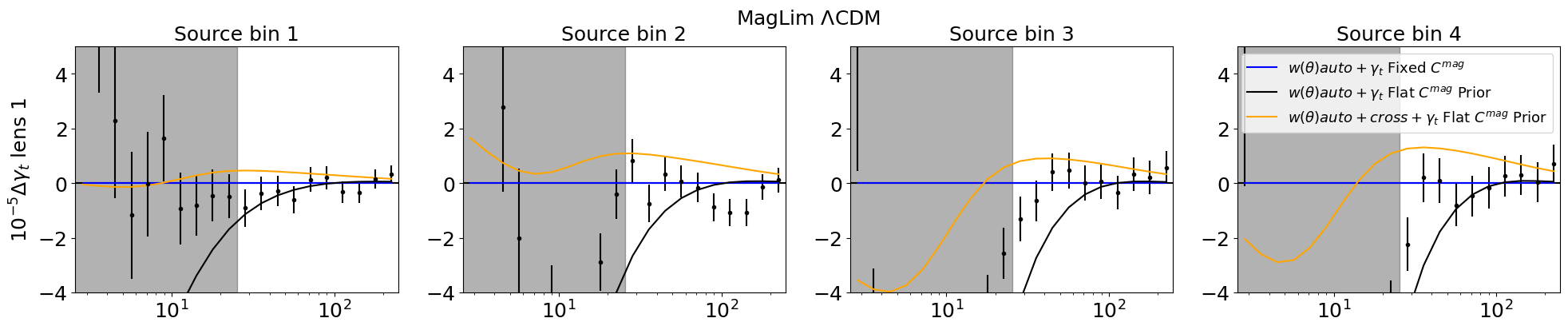}}
    \subfigure{\includegraphics[width=\linewidth]{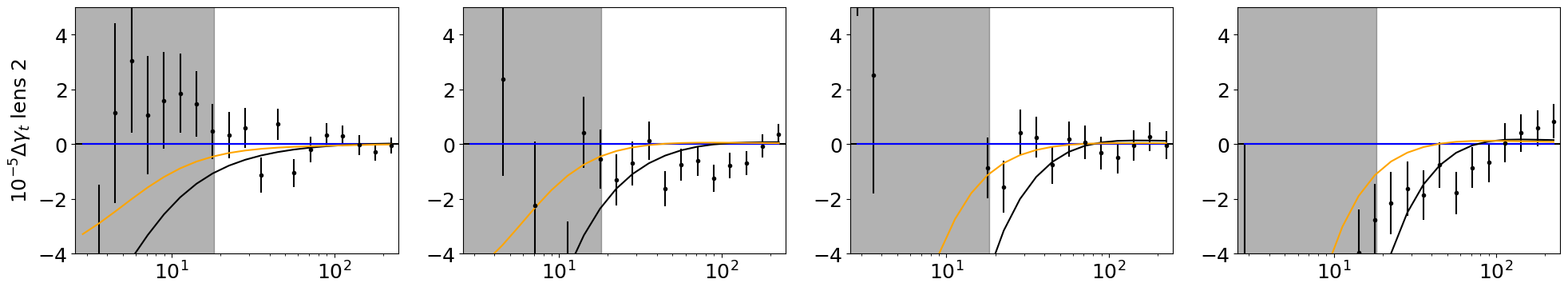}}
    \subfigure{\includegraphics[width=\linewidth]{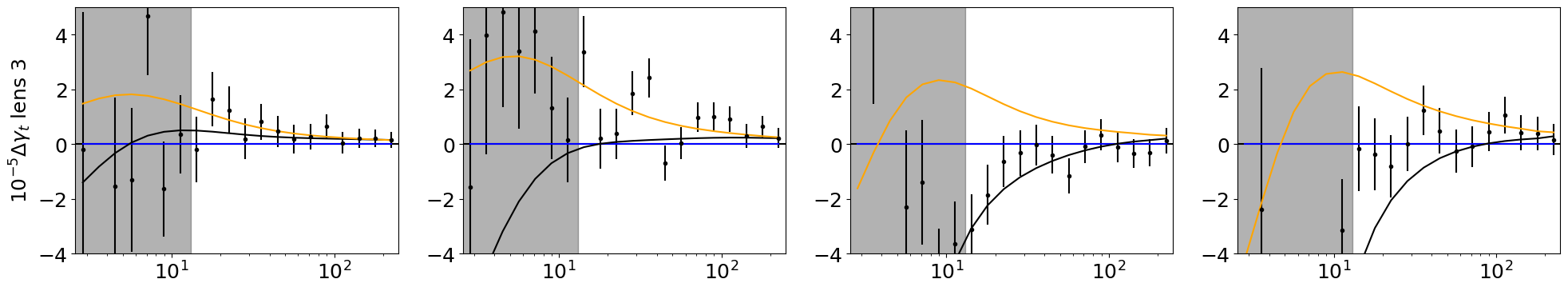}}
    \subfigure{\includegraphics[width=\linewidth]{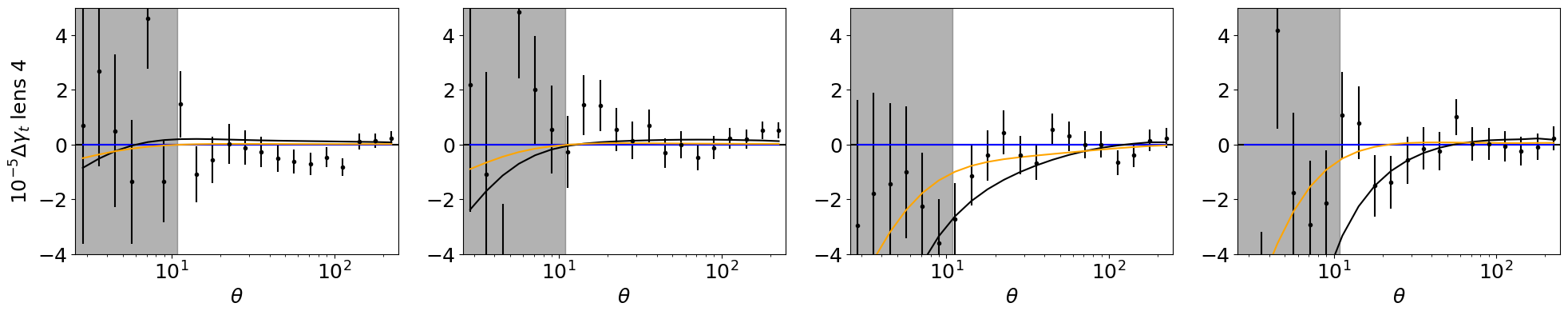}}
    
    \caption{ Residual $\gamma_t$ signal for the best fit of different magnification setups. }
    \label{fig:gt_residuals}
\end{figure*}

We also show the residual \maglim galaxy-galaxy lensing $\gamma_t$ signal for the best fit model from the fiducial fixed magnification chain in \fig{fig:gt_residuals}. It is of particular interest to look at the $\gamma_t$ signal for \maglim bin 3 as we find a $\csample^{3}$ posterior much higher than the fiducial \balrog value when using a flat magnification prior (as seen in \fig{fig:result_mag}). We find this signal decreases when magnification is freed, relative to the fixed case, due to the lower $\sigma_8$, despite the increase in the $\csample^{3}$ posterior. When the cross-lens bin clustering is added the the inference, the model best fit returns closer to the fixed case. 

\section{Statistical uncertainty in \Balrog}\label{app:stat_error}

In this appendix we derive the statistical uncertainty on the \balrog estimates in Equation \ref{eq:c_err_balrog}. 

The magnified and un-magnified \balrog samples contain a number of objects that are common between the two samples, we will call this \Nko, and a number of objects that are unique to each sample \Nkonly and \Noonly. We assume that these are three, independent Poisson distributed quantities with uncertainties

\begin{eqnarray}
    \sigma_{\Nko}^2 =& \Nko \\
    \sigma_{\Nkonly}^2 =& \Nkonly \\
    \sigma_{\Nkonly}^2 =& \Noonly
\end{eqnarray}

We can write the total number of objects in the two samples as, 

\begin{eqnarray}
    \Nk =& \Nkonly + \Nko \\
    \No =& \Noonly + \Nko 
\end{eqnarray}

Starting with Equation \ref{eq:c_estimate}, we can write \cbalrog as,

\begin{eqnarray}
    \cbalrog = \frac{\Nk - \No}{\No \dk} = \frac{X}{\No \dk},
\end{eqnarray}

and the uncertainty on this quantity as, 

\begin{eqnarray}
    \frac{\sigma_{\cbalrog}^2}{{\cbalrog}^2} = 
      \frac{\sigma_{X}^2}{X^2} 
    + \frac{\sigma_{\No}^2}{\No^2}
    - \frac{2 \sigma_{X\ \No}}{X\No}
\end{eqnarray}

We can then derive Equation \ref{eq:c_err_balrog} by substituting in the following relations,

\begin{eqnarray}
    X =& \Nkonly - \Noonly \\
    \sigma_X^2 =& \Nkonly + \Noonly  \\
    \sigma_{\No}^2 =& \No \\
    \sigma_{X\ \No} =& -\Noonly
\end{eqnarray}

The final statistical uncertainty is then, 

\begin{equation}
    \frac{\sigma^{\cbalrog}}{\cbalrog} = 
    \sqrt{\frac{ \Noonly + \Nkonly}{\left[\Nk - \No\right]^2} + \frac{1}{\No} + \frac{2\Noonly} {\No\left[\Nk - \No\right]}}.
\end{equation}

\end{document}